\documentclass[usenatbib,onecolumn]{mn2e}

\usepackage{epsf}

\newcommand\real{\mathrm{Re}}
\newcommand\imag{\mathrm{Im}}
\newcommand\rmD{\mathrm{D}}
\newcommand\rmd{\mathrm{d}}
\newcommand\rme{\mathrm{e}}
\newcommand\rmi{\mathrm{i}}

\newcommand\calE{\mathcal{E}}
\newcommand\calH{\mathcal{H}}
\newcommand\calI{\mathcal{I}}
\newcommand\f{\frac}
\newcommand\p{\partial}
\newcommand\cst{\mathrm{constant}}
\newcommand\sgn{\mathrm{sgn}}

\title[Satellite--disc interactions]
{Mean-motion resonances in satellite--disc interactions}

\author[Gordon I. Ogilvie]{Gordon I. Ogilvie\\
Department of Applied Mathematics and Theoretical Physics,
University of Cambridge, Centre for Mathematical Sciences,\\
Wilberforce Road, Cambridge CB3 0WA
}

\begin{document}

\maketitle

\label{firstpage}
 
\begin{abstract}
  Mean-motion resonances between a Keplerian disc and an orbiting
  companion are analysed within a Hamiltonian formulation using
  complex canonical Poincar\'e variables, which are ideally suited to
  the description of eccentricity and inclination dynamics.
  Irreversibility is introduced by allowing for dissipation within the
  disc.  A method is given for determining the rates of change of
  eccentricity and inclination variables of the disc and companion
  associated with resonances of various orders, including both
  reversible and irreversible effects, which extend and generalize
  previous results.  Preliminary applications to protoplanetary
  systems and close binary stars are discussed.
\end{abstract}

\begin{keywords}
  accretion, accretion discs --- binaries: close --- celestial mechanics ---
  planets: rings ---  planets and satellites: general
\end{keywords}

\section{Introduction}

\subsection{Satellite--disc interactions}

The gravitational interaction of a gaseous or particulate disc with an
orbiting companion is of general interest in astrophysics.  Important
examples include protoplanetary systems involving one or more planets
orbiting within a gaseous disc, accretion discs in close binary stars,
and systems of planetary rings and moons.

As in the planetary theory of classical celestial mechanics, the disc
and companion experience deviations from perfect Keplerian motion
around the central mass as a result of their mutual gravitational
perturbations.  Sufficiently close to the companion's orbit, the
behaviour of the disc can be complicated and highly nonlinear, leading
in many cases to an exclusion of the disc from the coorbital region.
The extent of such a gap depends on the ratio of the mass of the
companion to that of the central object, and on the properties of the
disc.  Away from this region, the interaction can be analysed by
perturbation methods and has two principal forms.

\emph{Secular interactions} (of first order in the perturbation) arise
from the mass distribution of the companion averaged around its orbit.
These contribute to the precessional behaviour of the system and allow
slow, reversible exchanges of eccentricity and inclination between the
companion and the disc.  Such behaviour is analogous to the
Laplace--Lagrange secular theory of planetary systems in which, for
example, Jupiter and Saturn undergo large-amplitude oscillatory
exchanges of eccentricity and inclination over tens of thousands of
years \citep[e.g.][]{MD99}.

\emph{Mean-motion resonances} depend specifically on the periodic
nature of the companion's orbit.  The fluctuating forces induced by
the companion generally give rise to rapid oscillations of small
amplitude in the disc that are of little consequence.  However, where
a commensurability occurs between the orbital periods of the companion
and a particle in the disc, a strong localized interaction occurs.
Through the intervention of collective effects and dissipation in the
disc, a different kind of secular behaviour is induced at second order
in the perturbation.  This leads to irreversible changes in semimajor
axis, eccentricity and inclination of the disc and companion, rather
than just precessional behaviour.

\subsection{Corotation and Lindblad resonances}

Mean-motion resonances in satellite--disc interactions were studied in
detail in an influential series of papers by Goldreich \& Tremaine.
Early in this series \citep{GT78,GT79,GT80} the authors considered a
two-dimensional problem in which a circular disc is perturbed by a
companion with a slightly eccentric orbit.  To calculate the linear
response of the disc, they decomposed the perturbing potential by
Fourier analysis into a series of rigidly rotating components with
various amplitudes, azimuthal wavenumbers and angular pattern speeds.
In a system with Keplerian orbits, the angular pattern speeds that
occur are rational multiples $p/q$ of the mean motion of the
companion, the amplitude scaling with the $|p-q|$th power of the
eccentricity.

The linearized equations governing the response of the disc at a given
azimuthal wavenumber and angular pattern speed allow for the
propagation of free waves in certain intervals of radius.  Three
important radii are the corotation radius, where the wave frequency as
measured in a frame rotating locally with the disc vanishes, and the
two Lindblad radii, where the same quantity equals (plus or minus) the
epicyclic frequency.  In the simplest case of a two-dimensional
non-self-gravitating disc, only one wave mode is permitted.  This
`density wave' combines inertial (epicyclic) and acoustic behaviour
and can propagate only where the wave frequency seen by the disc
exceeds the epicyclic frequency in magnitude.  It therefore propagates
interior to the inner Lindblad radius and exterior to the outer
Lindblad radius, but is evanescent between the Lindblad radii where
the corotation radius lies.

The forcing of disturbances by a perturbing potential component is
effectively localized to the neighbourhoods of the corotation and
Lindblad radii.  Beyond the Lindblad radii, the forcing is ineffective
because of the very limited overlap between the potential and the free
waves, which oscillate rapidly in radius.  Despite the evanescent
nature of the free waves near the corotation radius, it plays an
important role as a singularity of the linearized equations.  The
forcing can then be said to occur at corotation and Lindblad
\emph{resonances}, which can be identified with the mean-motion
resonances of celestial mechanics.  The reason for this is that, in a
Keplerian system where the epicyclic frequency equals the orbital
frequency, any radius where the orbital frequency is commensurate with
the mean motion of the companion serves, in principle, as both
corotation and Lindblad resonances for some potential components.

At a \emph{corotation resonance} the linearized response of the disc
at first order in the perturbation is singular but can be resolved by
including a viscosity.  At second order in the perturbation, the
companion exerts a secular torque on the disc as it generates a
localized, non-wavelike disturbance that transfers angular momentum
steadily to the disc through viscous stresses.

At a \emph{Lindblad resonance} the localized response is regularized
by collective effects such as pressure, self-gravity or viscosity, and
is generally in the form of an attenuated wave propagating away from
the resonance.  At second order in the perturbation, and provided that
the wave is dissipated somehow within the disc, the companion again
exerts a secular torque in launching the disturbance \citep[see
also][]{MS87}.

In a three-dimensional disc, not considered by \citet{GT79}, a
\emph{vertical resonance} is similar to a Lindblad resonance but
involves forcing and motion perpendicular to the plane of the disc, so
that a bending wave is excited rather than a density wave.

Convenient formulae are available for the corotation and Lindblad
torques, which have an important role in determining the orbital
evolution of the companion and the surface density distribution of the
disc.  The subsequent literature has made extensive use of these
expressions, as well as providing some modifications to the torque
formulae \citep{W89,A93,OL03}.  Together, these results provide the
basis of our understanding of planet--disc and other satellite--disc
interactions.

\subsection{Alternative treatments of mean-motion resonances}

It is perhaps not widely appreciated that the approach described above
is inadequate for general satellite--disc interactions.  A knowledge
of the resonant torque (by which is meant the component of the torque
perpendicular to the disc) determines one component of the rates of
change of the orbital angular momentum vectors of the disc and
companion and also, by virtue of Jacobi's theorem, the rate of change
of orbital energy of the companion; however, this information does not
generally suffice to determine the secular rates of change of the
relevant orbital elements of the disc and companion.  Furthermore, the
method does not apply to the general case in which the disc is
eccentric, because the forced wave equations in an eccentric disc have
not been analysed.

Later in their series of papers, \citet{GT81} and \citet{BGT84}
employed methods of celestial mechanics to treat cases in which either
the companion or the disc has an eccentric and/or inclined orbit.
Using `simple artifices' they derived formulae for the rates of change
of the canonical action variables associated with mean-motion
resonances, in terms of the classical disturbing function, while
avoiding an explicit discussion of collective effects.  In principle,
these expressions are considerably more powerful and general than the
earlier torque formulae, yet they appear not to have been used in the
subsequent literature.  We have tried to make use of them in
connection with eccentric planet--disc interactions and similar
problems but found that they generally give results that are not of
the desired form.

One aspect of their method is that it isolates, in the traditional
way, terms of different angular arguments in the disturbing function.
An implicit averaging is therefore involved over the mutual apsidal
and nodal precession of the disc and companion.  At least in some
contexts, this procedure is inappropriate because the mutual
precession may occur on a timescale comparable to that of the secular
evolution induced by mean-motion resonances, or indeed may not occur
at all if the disc and companion are engaged in a secular normal mode
\citep[e.g.][]{LO01}.  The formulae therefore describe the rates of
change of $a$, $e$ and $i$ averaged over the timescales of mutual
precession, while what are generally desired are the rates of change
of $a$, $e$, $i$, $\varpi$ and $\Omega$ averaged only over the
relative mean motion.\footnote{We use the standard notation for the
Keplerian orbital elements \citep[e.g.][]{MD99}: $a$, $e$, $i$,
$\varpi$ and $\Omega$ denote the semimajor axis, eccentricity,
inclination, longitude of pericentre and longitude of ascending node,
while $\lambda$ is the mean longitude.  The secular contribution to
the evolution of $\lambda$ is of less interest here.}

A second aspect is the way that dissipation is introduced into the
analysis, which is essential to capture the irreversible nature of the
effects of mean-motion resonances.  This is done either by including
ad hoc damping terms \citep{GT81} or is avoided altogether by using
Landau's prescription in which the disturbing function is given a slow
exponential growth \citep{BGT84}.  Although mathematically convenient,
these approaches can give misleading results.  The effects of
mean-motion resonances depend on the fact that the disc conserves
angular momentum while it dissipates energy, and the damping terms
added to the equations ought to respect these properties.  The Landau
approach is problematic because the first-order oscillatory
disturbances in the disc are permitted to grow exponentially rather
than being steadily dissipated, making it difficult to calculate the
second-order secular changes correctly.

The purpose of the present paper is to revisit the analysis of
mean-motion resonances in satellite--disc interactions and to derive
useful and general formulae for the associated evolution of orbital
elements of the disc and companion.  We employ methods of celestial
mechanics and treat the collective effects of the disc in an
approximate way that is adequate for our present purposes.  The
remainder of the paper is organized as follows.  In
Section~\ref{s:restricted} we consider the restricted problem in which
the companion has a prescribed orbit.  We develop the theory using a
Hamiltonian formulation of the equations, and describe the role of
collective effects.  In Section~\ref{s:unrestricted} we consider the
coupled satellite--disc system consistently and determine the feedback
of mean-motion resonances on the companion.  Preliminary applications
are worked out in Section~\ref{s:applications} and a summary of the
main ideas and results follows in Section~\ref{s:conclusions}.

\section{The restricted problem}
\label{s:restricted}

\subsection{Hamiltonian formulation}

We consider a thin disc around a central mass $M$, and initially
neglect any collective effects.  The disc is perturbed by an orbiting
companion of mass $M'$.  In the \emph{restricted problem} the
perturber has a prescribed periodic orbit that is unaffected by the
disc.  This approximation is usually applicable, for example, to
accretion discs in binary stars.  In this case the influence of the
companion on a test particle in the disc can be described through a
disturbing function that is a specified function of the position of
the particle and is also a periodic function of time.

In order to treat the dynamical equations in the most efficient way,
we adopt a Hamiltonian formulation using canonical variables
\citep[e.g.][]{M02}.  The mean longitude $\lambda$ is conjugate to the
energy-related action variable $\Lambda=(GMa)^{1/2}$, where $a$ is the
semimajor axis.  The other two generalized coordinates and conjugate
momenta, which relate to eccentricity and inclination and need not be
in action--angle form, are written as $q_\alpha$ and $p_\alpha$, with
$\alpha=\{1,2\}$.  Hamilton's equations for a test particle are
\begin{equation}
  \dot\lambda=\f{\p H}{\p\Lambda},\qquad
  \dot\Lambda=-\f{\p H}{\p\lambda},\qquad
  \dot q_\alpha=\f{\p H}{\p p_\alpha},\qquad
  \dot p_\alpha=-\f{\p H}{\p q_\alpha},
\end{equation}
where $H$ is the Hamiltonian per unit mass.\footnote{The true
Hamiltonian of the system is $\int H\,\rmd m$, where $\rmd m$ is a
mass element of the disc.} The unperturbed (Keplerian) Hamiltonian
\begin{equation}
  H_0=-\f{(GM)^2}{2\Lambda^2}=-\f{GM}{2a}
\end{equation}
gives rise to a mean motion
\begin{equation}
  \dot\lambda=n(\Lambda)=\f{\rmd H_0}{\rmd\Lambda}=\f{(GM)^2}{\Lambda^3}=
  \left(\f{GM}{a^3}\right)^{1/2}.
\end{equation}

The disturbing function is periodic in the mean longitudes $\lambda$
and $\lambda'$ of the test particle and the perturber, and can
therefore be expanded in a double Fourier series.  We consider a
Hamiltonian of the form
\begin{equation}
  H=H_0(\Lambda)-\epsilon\,\real\left[
  R(\Lambda,q_\alpha,p_\alpha)\,\rme^{\rmi\phi}\right],
\label{h}
\end{equation}
where $\epsilon=1$ is a bookkeeping parameter used to identify effects
of various orders in the perturbation, $R$ is a complex potential
amplitude (proportional to $M'$) and $\phi=k\lambda+k'\lambda'$ is a
potentially resonant angle with integer coefficients $k$ and
$k'$.\footnote{Note that it is permissible to replace $R$ with $R^*$
if the signs of $k$ and $k'$ are both changed.}  In the restricted
problem $\lambda'=n't+\epsilon'$ is a prescribed linear function of
time.  The double Fourier series in fact gives rise to an infinite
number of perturbing terms of this form, which might be labelled using
the notation $R_{k,k'}$ and $\phi_{k,k'}$.  We may consider them
separately because their second-order secular effects will sum in
quadrature owing to the phase relations between them.  Hamilton's
equations are then
\begin{equation}
  \dot\lambda=n(\Lambda)-
  \epsilon\,\real\left[\f{\p R}{\p\Lambda}\,\rme^{\rmi\phi}\right],
\end{equation}
\begin{equation}
  \dot\Lambda=\epsilon\,\real\left[\rmi k R\,\rme^{\rmi\phi}\right],
\end{equation}
\begin{equation}
  \dot q_\alpha=-\epsilon\,\real\left[\f{\p R}{\p p_\alpha}\,
  \rme^{\rmi\phi}\right],
\end{equation}
\begin{equation}
  \dot p_\alpha=\epsilon\,\real\left[\f{\p R}{\p q_\alpha}\,
  \rme^{\rmi\phi}\right].
\end{equation}
An important point is that the apsidal and nodal precessional
longitudes $\varpi$ and $\Omega$ appear in $R$ (through its dependence
on $q_\alpha$ and $p_\alpha$) and not in the exponential.  Terms in
the disturbing function with the same values of $k$ and $k'$ but
different dependences on $\varpi$ and $\Omega$ are treated together,
not separately.  To isolate such terms would be appropriate only if we
were to average the behaviour of the system over the mutual apsidal
and nodal precession of the disc and companion.

\subsection{Perturbative expansion to second order}
\label{s:perturbative}

We expand the dynamical variables in the form
$\Lambda=\Lambda_0+\epsilon\Lambda_1+\epsilon^2\Lambda_2+\cdots$,
etc., with the intention that $\Lambda_0$ describes the unperturbed
disc, $\Lambda_1$ the periodic oscillation induced by the perturbation
at first order, and $\Lambda_2$ contains second-order variations.  The
effects that we are looking for will appear as secular changes of
$\Lambda_2$, etc.\footnote{This form of perturbative expansion is
valid only for a limited time interval, because $\Lambda_2$ will
eventually become comparable to $\Lambda_0$.  A more sophisticated
approach would be to use the method of multiple timescales, which
would yield a uniformly asymptotic solution.  However, the
notationally simpler approach taken here allows us to calculate the
secular rates of change correctly.}

At leading order all variables are independent of time except for
\begin{equation}
  \dot\lambda_0=n(\Lambda_0)=n_0,
\end{equation}
which describes the unperturbed mean motion of the disc.  It is
convenient to use either the unperturbed semimajor axis $a_0$ or the
related action variable $\Lambda_0=(GMa_0)^{1/2}$ to label the orbits
in the disc.  Then
\begin{equation}
  \dot\phi_0=kn_0+k'n'
\end{equation}
is a constant depending on $\Lambda_0$.  By definition, it vanishes at
the location of the mean-motion resonance where $n_0/n'=-k'/k$ is a
certain positive rational number.

At first order we find from Hamilton's equations
\begin{equation}
  \dot\lambda_1=\f{\rmd n}{\rmd\Lambda}\Lambda_1-
  \real\left[\f{\p R}{\p\Lambda}\,\rme^{\rmi\phi_0}\right],
\label{dotl1}
\end{equation}
\begin{equation}
  \dot\Lambda_1=\real\left[\rmi k R\,\rme^{\rmi\phi_0}\right],
\label{dotL1}
\end{equation}
\begin{equation}
  \dot q_{\alpha1}=-\real\left[\f{\p R}{\p p_\alpha}\,\rme^{\rmi\phi_0}\right],
\label{dotq1}
\end{equation}
\begin{equation}
  \dot p_{\alpha1}=\real\left[\f{\p R}{\p q_\alpha}\,\rme^{\rmi\phi_0}\right],
\label{dotp1}
\end{equation}
where all terms on the right-hand sides are evaluated on the
unperturbed solution, i.e. at $\Lambda=\Lambda_0$,
$q_\alpha=q_{\alpha0}$, $p_\alpha=p_{\alpha0}$.  The solution of
equations (\ref{dotL1})--(\ref{dotp1}) is
\begin{equation}
  \Lambda_1=\real\left[fkR\,\rme^{\rmi\phi_0}\right],
\end{equation}
\begin{equation}
  q_{\alpha1}=\real\left[\rmi f\f{\p R}{\p p_\alpha}\,\rme^{\rmi\phi_0}\right],
\end{equation}
\begin{equation}
  p_{\alpha1}=\real\left[-\rmi f\f{\p R}{\p q_\alpha}\,
  \rme^{\rmi\phi_0}\right],
\end{equation}
where
\begin{equation}
  f=\f{1}{\dot\phi_0}.
\label{f}
\end{equation}
Equation (\ref{dotl1}) then becomes
\begin{equation}
  \dot\lambda_1=\real\left[\f{\rmd n}{\rmd\Lambda}fkR\,
  \rme^{\rmi\phi_0}-\f{\p R}{\p\Lambda}\,\rme^{\rmi\phi_0}\right]=
  \real\left[-f^{-1}\f{\p}{\p\Lambda}(fR)\,\rme^{\rmi\phi_0}\right],
\end{equation}
for which the solution is
\begin{equation}
  \lambda_1=\real\left[\rmi\f{\p}{\p\Lambda}(fR)\,
  \rme^{\rmi\phi_0}\right],
\end{equation}
and then we find $\phi_1=k\lambda_1$.  To this order, therefore, the
variables undergo forced oscillations around their leading-order
values with angular frequency $\dot\phi_0$.  Four arbitrary constants
could be included in the solution, corresponding to small shifts in
the mean values of the variables, but these are best absorbed into the
leading-order terms.  The amplitude of the oscillations diverges at
the resonance where $\dot\phi_0=0$.

At second order the equation for $\dot\lambda_2$ is not required,
but we have also
\begin{equation}
  \dot\Lambda_2=\real\bigg[\rmi k\bigg(\f{\p R}{\p\Lambda}\Lambda_1+
  \f{\p R}{\p q_\beta}q_{\beta1}+\f{\p R}{\p p_\beta}p_{\beta1}+
  \rmi\phi_1R\bigg)\rme^{\rmi\phi_0}\bigg],
\end{equation}
\begin{equation}
  \dot q_{\alpha2}=-\real\bigg[\bigg(\f{\p^2R}{\p p_\alpha\p\Lambda}\Lambda_1+
  \f{\p^2R}{\p p_\alpha\p q_\beta}q_{\beta1}+\f{\p^2R}{\p p_\alpha\p p_\beta}
  p_{\beta1}+\rmi\phi_1\f{\p R}{\p p_\alpha}\bigg)\rme^{\rmi\phi_0}\bigg],
\end{equation}
\begin{equation}
  \dot p_{\alpha2}=\real\bigg[\bigg(\f{\p^2R}{\p q_\alpha\p\Lambda}\Lambda_1+
  \f{\p^2R}{\p q_\alpha\p q_\beta}q_{\beta1}+\f{\p^2R}{\p q_\alpha\p p_\beta}
  p_{\beta1}+\rmi\phi_1\f{\p R}{\p q_\alpha}\bigg)\rme^{\rmi\phi_0}\bigg],
\end{equation}
where, again, terms on the right-hand sides are evaluated on the
unperturbed solution, and we adopt the convention that \emph{a
summation over $\beta=\{1,2\}$ will be implied wherever it appears}.
The right-hand sides, involving products of oscillating quantities,
contain both non-oscillatory terms of zero frequency and oscillatory
terms of frequency $2\dot\phi_0$.  We are interested in the secular
rates of change of $\Lambda$, $q_\alpha$ and $p_\alpha$ at second
order in $\epsilon$, which we denote by angle brackets.  Making use of
the relation
\begin{equation}
  \real\left[A\,\rme^{\rmi\phi_0}\,
  \real\left[B\,\rme^{\rmi\phi_0}\right]\right]=
  \f{1}{2}\real\left[AB\,\rme^{2\rmi\phi_0}+A^*B\right]
\end{equation}
and discarding the oscillatory terms, we find
\begin{equation}
  \langle\dot\Lambda\rangle=\f{1}{2}\,\real\bigg[
  fk\left(\f{\p R^*}{\p q_\beta}\f{\p R}{\p p_\beta}-
  \f{\p R^*}{\p p_\beta}\f{\p R}{\p q_\beta}\right)-
  \rmi k^2\f{\p}{\p\Lambda}(f|R|^2)\bigg],
\label{secular1}
\end{equation}
\begin{equation}
  \langle\dot q_\alpha\rangle=-\f{1}{2}\,\real\bigg[
  \rmi f
  \left(\f{\p^2R^*}{\p p_\alpha\p q_\beta}\f{\p R}{\p p_\beta}-
  \f{\p^2R^*}{\p p_\alpha\p p_\beta}\f{\p R}{\p q_\beta}\right)+
  k\f{\p}{\p\Lambda}\left(fR\f{\p R^*}{\p p_\alpha}
  \right)\bigg],
\label{secular2}
\end{equation}
\begin{equation}
  \langle\dot p_\alpha\rangle=\f{1}{2}\,\real\bigg[
  \rmi f
  \left(\f{\p^2R^*}{\p q_\alpha\p q_\beta}\f{\p R}{\p p_\beta}-
  \f{\p^2R^*}{\p q_\alpha\p p_\beta}\f{\p R}{\p q_\beta}\right)+
  k\f{\p}{\p\Lambda}\left(fR\f{\p R^*}{\p q_\alpha}
  \right)\bigg].
\label{secular3}
\end{equation}

So far we have not used the fact that $f$ is a real quantity; the
reason for this will become apparent later.  Away from the resonance,
$f$ is real and finite and equations
(\ref{secular1})--(\ref{secular3}) become
\begin{equation}
  \langle\dot\Lambda\rangle=\f{\p\calH}{\p\lambda}=0,
\end{equation}
\begin{equation}
  \langle\dot q_\alpha\rangle=\f{\p\calH}{\p p_\alpha},
\end{equation}
\begin{equation}
  \langle\dot p_\alpha\rangle=-\f{\p\calH}{\p q_\alpha},
\end{equation}
where
\begin{equation}
  \calH=\f{1}{4}\left[
  -\rmi f\left(\f{\p R^*}{\p q_\beta}\f{\p R}{\p p_\beta}-
  \f{\p R^*}{\p p_\beta}\f{\p R}{\p q_\beta}\right)-
  k\f{\p}{\p\Lambda}(f|R|^2)\right]
\label{calH}
\end{equation}
is a real Hamiltonian.

The implication of our analysis is that each oscillatory term
$R_{k,k'}$ in the disturbing function gives rise to an oscillatory
behaviour in the disc at first order and to a secular behaviour at
second order.  Since it is described by a Hamiltonian $\calH_{k,k'}$
that is independent of $\lambda$, this secular behaviour concerns only
the eccentricity and inclination variables and is reversible and
precessional in character.  It is of interest mainly because it
diverges in the vicinity of mean-motion resonances where $f\to\infty$.
The resolution of this singularity, which results in non-Hamiltonian,
irreversible secular behaviour, is discussed in
Section~\ref{s:resolution}.

The above procedure does not work for the secular part $R_{0,0}$ of
the disturbing function, for which $\dot\phi_0$ vanishes identically.
However, this term contributes directly to the secular dynamics at
first order in the perturbation, giving rise to the familiar
precessional effects, and we are not concerned with its smaller,
second-order counterpart.  Second-order effects are of interest only
because of their amplification by mean-motion resonances.  The secular
dynamics is therefore adequately described by a Hamiltonian consisting
of $-R_{0,0}$ together with the sum of $\calH_{k,k'}$ for any
important resonances $\{k,k'\}$, subject to the resolution of the
corresponding singularities.

\subsection{Complex canonical variables}
\label{s:complex}

Although Hamilton's equations employ real variables, complex variables
are natural in celestial mechanics, especially for describing
precessional behaviour.  The idea of complex canonical variables was
employed by \citet{S66} in order to unify the descriptions of
classical and quantum-mechanical systems.  Suppose we have a
Hamiltonian system with canonical variables $q_\alpha$ and $p_\alpha$
and Hamiltonian $H(q_\alpha,p_\alpha)$.\footnote{In this paragraph we
assume that $\alpha$ labels all the degrees of freedom, not just a
subset.}  Then $(q_\alpha,p_\alpha)$ can be replaced by
$(z_\alpha,z_\alpha^*)$, where
\begin{equation}
  z_\alpha=\f{1}{\sqrt{2}}(q_\alpha+\rmi p_\alpha).
\end{equation}
(In order for the dimensions to match, $q_\alpha$ and $p_\alpha$
should be variables of `rectangular' rather than `action--angle' form.)
Hamilton's equations then combine in the compact form
\begin{equation}
  \rmi\dot z_\alpha=\f{\p H}{\p z_\alpha^*},
\label{strocchi}
\end{equation}
in which $H$ is now regarded as a real-valued analytic function of
$z_\alpha$ and $z_\alpha^*$, treated as algebraically independent
variables.  (The conjugate equation $-\rmi\dot z_\alpha^*=\p H/\p
z_\alpha$ need not be considered separately, as it follows directly
from the complex conjugate of equation~(\ref{strocchi}); $z_\alpha$
and $z_\alpha^*$ are `conjugate' variables in both senses of the
word.)

The equations derived so far in this paper are valid for any choice of
canonical eccentricity and inclination variables $q_\alpha$ and
$p_\alpha$ used in conjunction with $\lambda$ and $\Lambda$, such as
modified Delaunay variables \citep[e.g.][]{M02}.  Indeed the quantity
\begin{equation}
  \f{\p R^*}{\p q_\beta}\f{\p R}{\p p_\beta}-
  \f{\p R^*}{\p p_\beta}\f{\p R}{\p q_\beta}=\{R^*,R\}
\label{pb}
\end{equation}
that appears in equation (\ref{calH}) is a Poisson bracket and is
therefore invariant under canonical transformations.  (The variables
$\lambda$ and $\Lambda$ do not appear in this Poisson bracket because
$R$ is independent of $\lambda$.)  We will use the canonical variables
\begin{equation}
  q_1=\left\{2\Lambda\left[1-(1-e^2)^{1/2}\right]\right\}^{1/2}\cos\varpi,
\label{q1}
\end{equation}
\begin{equation}
  p_1=\left\{2\Lambda\left[1-(1-e^2)^{1/2}\right]\right\}^{1/2}\sin\varpi,
\end{equation}
\begin{equation}
  q_2=\left[2\Lambda(1-e^2)^{1/2}(1-\cos i)\right]^{1/2}\cos\Omega,
\end{equation}
\begin{equation}
  p_2=\left[2\Lambda(1-e^2)^{1/2}(1-\cos i)\right]^{1/2}\sin\Omega,
\label{p2}
\end{equation}
which are identical to the rectangular variables of \citet{P92}, but
rotated through an angle of $\pi/2$.  When $e\ll1$ and $i\ll1$, these
variables reduce to
$(q_1,p_1)\approx\Lambda^{1/2}(e\cos\varpi,e\sin\varpi)$ and
$(q_2,p_2)\approx\Lambda^{1/2}(i\cos\Omega,i\sin\Omega)$, showing
their relation to the Cartesian components of the eccentricity and
inclination vectors \citep[e.g.][]{MD99}.

Strocchi's transformation leads us to consider
\begin{equation}
  z_1=\left\{\Lambda\left[1-(1-e^2)^{1/2}\right]\right\}^{1/2}
  \,\rme^{\rmi\varpi},
\end{equation}
\begin{equation}
  z_2=\left[\Lambda(1-e^2)^{1/2}(1-\cos i)\right]^{1/2}\,\rme^{\rmi\Omega},
\end{equation}
which may be called \emph{complex canonical Poincar\'e variables} and
were also employed by \citet{LR95}.\footnote{They can alternatively be
obtained from the modified Delaunay variables via the transformation
$z_\alpha=p_\alpha^{1/2}\,\rme^{-\rmi q_\alpha}$.}
Note that
\begin{equation}
  L=\Lambda-|z_1|^2=[GMa(1-e^2)]^{1/2}
\end{equation}
is the total specific angular momentum and
\begin{equation}
  L_z=\Lambda-|z_1|^2-|z_2|^2=[GMa(1-e^2)]^{1/2}\cos i
\label{lz}
\end{equation}
is its component perpendicular to the reference plane.  Since
$\Lambda$ is the specific angular momentum of a circular orbit of
radius $a$ in the reference plane, $|z_1|^2+|z_2|^2$ is the specific
\emph{angular momentum deficit} or AMD \citep{L97}.  It consists of
two parts, $|z_1|^2$ associated with eccentricity and $|z_2|^2$ with
inclination, and is a positive-definite measure of the deviation of an
orbit from circularity and coplanarity (with respect to the arbitrary
reference plane).

We now consider the complex potential amplitude $R$ as an analytic
function of $\Lambda$, $z_\alpha$ and $z_\alpha^*$ satisfying the
analytic properties
\begin{equation}
  \left(\f{\p R}{\p z_\alpha}\right)^*=\f{\p R^*}{\p z_\alpha^*},\qquad
  \left(\f{\p R}{\p z_\alpha^*}\right)^*=\f{\p R^*}{\p z_\alpha}.
\end{equation}
The Poisson bracket of equation (\ref{pb}) becomes
\begin{equation}
  \{R^*,R\}=-\rmi\left(\f{\p R^*}{\p z_\beta}\f{\p R}{\p z_\beta^*}-
  \f{\p R^*}{\p z_\beta^*}\f{\p R}{\p z_\beta}\right)
\end{equation}
and so equation (\ref{calH}) for the second-order Hamiltonian reads
\begin{equation}
  \calH=\f{1}{4}\left[f
  \left(\left|\f{\p R}{\p z_\beta}\right|^2-
  \left|\f{\p R}{\p z_\beta^*}\right|^2\right)-
  k\f{\p}{\p\Lambda}(f|R|^2)\right].
\label{hz}
\end{equation}
It will be seen below that the terms in the first bracket, which arise
from the dynamics of the eccentricity and inclination variables
$z_\alpha$, correspond in some sense to Lindblad and vertical
resonances, while the term involving a derivative with respect to
$\Lambda$, which arises from the dynamics of the mean-motion variables
$\Lambda$ and $\lambda$, corresponds in some sense to corotation
resonances.  A similar distinction is apparent in the analysis of
\citet{BGT84}.  We therefore refer to the first and second parts as
the \emph{Lindblad/vertical term} and the \emph{corotation term},
respectively.

\subsection{The disturbing function in complex canonical Poincar\'e variables}

The disturbing function is usually expanded in the variables $e$ and
$s=\sin(i/2)$.  In converting these into complex Poincar\'e variables
it is useful to define the dimensionless complex eccentricity and
inclination variables
\begin{equation}
  \calE=\left(\f{2}{\Lambda}\right)^{1/2}z_1,
\end{equation}
\begin{equation}
  \calI=\left(\f{2}{\Lambda}\right)^{1/2}z_2,
\end{equation}
and then to note that
\begin{equation}
  e\,\rme^{\rmi\varpi}=\calE\left(1-{\textstyle\f{1}{4}}|\calE|^2\right)^{1/2},
\end{equation}
\begin{equation}
  2s\,\rme^{\rmi\Omega}=\calI
  \left(1-{\textstyle\f{1}{2}}|\calE|^2\right)^{-1/2}.
\end{equation}
Therefore the magnitudes of $\calE$ and $\calI$ are directly related
to the eccentricity and inclination and are approximately equal to
them when small, while the phases of $\calE$ and $\calI$ are the
corresponding precessional longitudes $\varpi$ and $\Omega$.

Expansions for $R$ in terms of $\calE$ and $\calI$ up to fourth degree
in eccentricity and inclination are given for the most important
commensurabilities in Appendix~\ref{s:df}.  \citet{LR95} provide a
method for generating such expansions, and prove theorems regarding
their general algebraic form.  In fact, the expressions in
Appendix~\ref{s:df} were obtained simply by rewriting the expansion of
\citet{MD99} in terms of $\calE$ and $\calI$.  (Note, however, that
primed variables always refer here to the companion, regardless of
whether the resonance lies interior or exterior to it.)

For the purposes of illustration, consider a first-order interior
$j:j-1$ resonance with $j\ge2$.  By this notation we mean that the
resonant orbit in the disc is interior to the perturber and has a mean
motion $j/(j-1)$ times that of the companion.  We define the ratio of
semimajor axes $\alpha=a/a'=[(j-1)/j]^{2/3}$.  \citet{MD99} give the
disturbing function as the sum of direct and indirect terms in the form
\begin{equation}
  {\cal R}=\f{GM'}{a'}\left({\cal R}_\mathrm{D}+
  \alpha{\cal R}_\mathrm{E}\right).
\end{equation}
The terms of lowest degree in eccentricity and inclination for this
commensurability are \citep[][Tables~B.4 and~B.5]{MD99}
\begin{equation}
  {\cal R}_\mathrm{D}=ef_{27}\cos[j\lambda'-(j-1)\lambda-\varpi]+
  e'f_{31}\cos[j\lambda'-(j-1)\lambda-\varpi'],
\end{equation}
\begin{equation}
  {\cal R}_\mathrm{E}=-2e'\cos[j\lambda'-(j-1)\lambda-\varpi']\delta_{j,2},
\end{equation}
where $\delta$ is the Kronecker delta and
\begin{equation}
  f_{27}=\f{1}{2}\left(-2j-\alpha\f{\rmd}{\rmd\alpha}\right)
  b_{1/2}^{(j)}(\alpha),
\end{equation}
\begin{equation}
  f_{31}=\f{1}{2}\left(-1+2j+\alpha\f{\rmd}{\rmd\alpha}\right)
  b_{1/2}^{(j-1)}(\alpha)
\end{equation}
are given in terms of Laplace coefficients.  Comparing with
equation~(\ref{h}), we may therefore identify the integer coefficients
$k=j-1$ and $k'=-j$, the corresponding frequency
$\dot\phi_0=(j-1)n-jn'$ and the complex potential amplitude
\begin{equation}
  R\approx\f{GM'}{a'}\left[f_{27}\calE+
  (f_{31}-2\alpha\delta_{j,2})\calE'\right].
\end{equation}
The equivalent expression correct to fourth degree in eccentricity and
inclination is given in equation~(\ref{first-interior}), which
compactly combines all terms in Tables~B.4 and~B.5 of \citet{MD99}.
Note that the inclination terms vanish when $\calI=\calI'$, i.e.~when
the disc and perturber are coplanar.

\subsection{Resolution of the resonant singularity}
\label{s:resolution}

As a simple but important example, at a second-order interior
resonance with a perturber having a circular and coplanar orbit, such
as the $3:1$ resonance that occurs in accretion discs in close binary
stars of sufficiently small mass ratio, the dominant term in the
disturbing function has $R\propto\calE^2$ for $e\ll1$ and is therefore
of the form
\begin{equation}
  R=g(\Lambda)z_1^2+O(z^4),
\end{equation}
where, in fact, $g$ is real (cf.~equation~\ref{second-interior}).
According to equation (\ref{hz}), the second-order Hamiltonian arising
from this disturbing function is
\begin{equation}
  \calH=\dot\phi_0^{-1}g^2|z_1|^2+O(z^4),
\end{equation}
which gives rise to the dynamics
\begin{equation}
  \rmi\langle\dot z_1\rangle=\f{\p\calH}{\p z_1^*}=
  \dot\phi_0^{-1}g^2z_1+O(z^3).
\label{iz1dot}
\end{equation}
Since $\arg(z_1)=\varpi$, this behaviour corresponds simply to apsidal
precession at a rate $-g^2/\dot\phi_0$, which is retrograde interior
to the resonance (where $\dot\phi_0>0$), prograde exterior to the
resonance (where $\dot\phi_0<0$) and diverges at the resonance itself.
This behaviour can be seen, for example, in Fig.~1 of \citet{KMFFF91},
which shows how the mean apsidal and nodal precession rates of
asteroids in nearly circular orbits, calculated to second order in the
planetary masses, are affected in the vicinity of mean-motion
resonances with Jupiter.  The $3:1$ and $5:3$ resonances display the
characteristic behaviour described here for second-order interior
resonances.

This divergence arises because the expansion we have adopted breaks
down in the vicinity of the resonance.  The behaviour of
non-interacting particles and of continuous discs near a mean-motion
resonance is different.  Single particles avoid the divergence by
undergoing libration, which means that the resonant angle, instead of
circulating arbitrarily slowly, oscillates about an equilibrium value.
Fluid or collisional particle discs, in which the intersection of
streamlines is resisted, can avoid the divergence through the
intervention of collective effects such as pressure, self-gravity or
viscosity.  \citet{MS87} showed how any of these collective effects,
in the case of a Lindblad resonance, acts to displace the resonant
pole slightly away from the real axis, after a spatial Fourier
transform of the linearized perturbation equations is carried out.  We
initially adopt a similar prescription, replacing equation~(\ref{f})
with
\begin{equation}
  f=\f{1}{\dot\phi_0-\rmi s},
\label{replace}
\end{equation}
where $s$ is a positive parameter with the dimensions of frequency.
This replacement can also be motivated by Landau's approach, in which
the periodic disturbing function is `turned on' slowly by including an
additional time-dependence proportional to $\rme^{st}$
\citep[cf.][]{BGT84}.  However, Landau's approach must be used with
care and we return to address this issue more thoroughly in
Section~\ref{s:collective}.

When $f$ is treated throughout the analysis of
Section~\ref{s:perturbative} as a complex quantity, we obtain secular
evolutionary equations at second order that are not of Hamiltonian
form but allow for the irreversibility that is associated with the
resonance.  In particular,
\begin{equation}
  \langle\dot\Lambda\rangle=-\f{1}{2}kf_\rmi
  \left(\left|\f{\p R}{\p z_\beta}\right|^2-
  \left|\f{\p R}{\p z_\beta^*}\right|^2\right)+
  \f{1}{2}k^2\f{\p}{\p\Lambda}(f_\rmi|R|^2),
\end{equation}
\begin{equation}
  \rmi\langle\dot z_\alpha\rangle=\f{1}{4}f\left(\f{\p^2R^*}
  {\p z_\alpha^*\p z_\beta^*}\f{\p R}{\p z_\beta}-
  \f{\p^2R^*}{\p z_\alpha^*\p z_\beta}\f{\p R}{\p z_\beta^*}\right)+
  \f{1}{4}f^*\left(\f{\p^2R}{\p z_\alpha^*\p z_\beta}\f{\p R^*}{\p z_\beta^*}-
  \f{\p^2R}{\p z_\alpha^*\p z_\beta^*}\f{\p R^*}{\p z_\beta}\right)-
  \f{1}{4}k\f{\p}{\p\Lambda}\left(fR\f{\p R^*}{\p z_\alpha^*}+
  f^*R^*\f{\p R}{\p z_\alpha^*}\right),
\end{equation}
where $f_\rmi=\imag[f]$; these expressions agree with
$\p\calH/\p\lambda$ and $\p\calH/\p z_\alpha^*$ if $f_\rmi=0$.
In the present example we obtain, for $e\ll1$,
\begin{equation}
  \rmi\langle\dot z_1\rangle=fg^2z_1=
  \left(\f{\dot\phi_0+\rmi s}{\dot\phi_0^2+s^2}\right)g^2z_1,
\end{equation}
instead of equation~(\ref{iz1dot}).  Resolution of the resonant
singularity therefore leads to a finite precession rate of
$-g^2\dot\phi_0/(\dot\phi_0^2+s^2)$ together with a qualitatively
different effect: an eccentricity growth rate of
$g^2s/(\dot\phi_0^2+s^2)$.  The growth rate peaks at the resonance,
the height and width of the peak depending on the parameter $s$, while
the integrated growth rate is independent of $s$.  In fact the growth
rate can be represented as $\pi g^2\delta(\dot\phi_0)$ as $s\to0$,
i.e.~in the case of a resonance that is not significantly broadened.
We note also that
\begin{equation}
  \delta(\dot\phi_0)=\left|\f{\rmd\dot\phi_0}{\rmd\Lambda_0}\right|^{-1}
  \delta(\Lambda_0-\hat\Lambda_0)=
  \left|\f{\rmd\dot\phi_0}{\rmd a_0}\right|^{-1}\delta(a_0-\hat a_0),
\end{equation}
where $\hat\Lambda_0$ (or $\hat a_0$) locates the resonant orbit.
Unlike the Hamiltonian precessional behaviour, the eccentricity growth
is irreversible and corresponds to a growth of the AMD of the disc:
\begin{equation}
  \bigg\langle\f{\rmd|z_1|^2}{\rmd t}\bigg\rangle=
  \left(\f{2s}{\dot\phi_0^2+s^2}\right)g^2|z_1|^2\sim2\pi g^2|z_1|^2\,
  \delta(\dot\phi_0).
\end{equation}

The physical resolution of the divergent precession rate is therefore
a finite precession rate together with an irreversible growth or decay
(in this case a growth of eccentricity).  This behaviour is
characteristic of an eccentric Lindblad resonance as described by
\citet{L91}.  For the $3:1$ resonance in a system with mass ratio
$q=M'/M\ll1$ we obtain $g\approx1.727\,q(GM/a'^3)^{1/2}$ and an
eccentricity growth rate of
$2.082\,q^2(GM/a'^3)^{1/2}a_0\,\delta(a_0-\hat a_0)$.  This agrees
with the result of \citet{L91} as interpreted by \citet{GO06}.
However, neither of these papers considered the second-order
precessional effect.

Applying equation~(14) of \citet{BGT84}, however, gives exactly
\emph{twice} the eccentricity growth rate of \citet{L91}.  The reason
for this is that their (modified Delaunay) variable $\Gamma$ is
proportional to $e^2$ for $e\ll1$.  In their analysis $e^2$ grows
partly because the non-oscillatory part of $e$ grows secularly, which
is physically correct, but also because the oscillatory part of $e$
grows proportionally to $\rme^{st}$.  In reality the oscillatory part
of $e$ has a steady amplitude owing to dissipation in the disc.
Therefore application of Landau's prescription in this case gives a
misleading result for the second-order behaviour of the disc.

\subsection{Collective effects in the disc}
\label{s:collective}

We now attempt to give a justification for the procedure described
above.  The detailed behaviour of the disc near a resonance depends,
in principle, on a number of factors: the relative importance of
various collective effects (such as pressure, self-gravity or
viscosity), the level of nonlinearity, the vertical structure of a
three-dimensional disc, etc.  It is known that the torque exerted at a
Lindblad resonance is independent of these details provided that the
response is linear and localized near the resonant orbit
\citep{MS87,LO98}, while nonlinearity in fact makes little difference
\citep{YC94}.  On the other hand, the torque exerted at a corotation
resonance does depend in an important way on the relative importance
of nonlinearity and viscosity \citep{OL03}.

We therefore adopt a minimal description of the collective effects in
the disc, by introducing viscous behaviour without attempting the
formidable task of expressing the full Navier--Stokes equation in
terms of orbital elements.  The dynamical equations adopted so far
refer to individual particles, and the orbital elements of a given
particle are functions of time only.  To deal with a continuous disc
we adopt a semi-Lagrangian description with spatial coordinates
$(a_0,\lambda_0)$ labelling the fluid elements at any
instant.\footnote{For a three-dimensional description a third
coordinate is required to describe the distance from the `midplane' of
the disc.}  Note that the unperturbed variables satisfy $a_0=\cst$ and
$\lambda_0-n_0t=\epsilon_0=\cst$ for any fluid element, with
$n_0=(GM/a_0^3)^{1/2}$.  Therefore $(a_0,\epsilon_0)$ are true
Lagrangian coordinates but are less suitable because of the rapid
shearing in the unperturbed flow.  The total time-derivative
$\rmd/\rmd t$ translates into the Lagrangian derivative
\begin{equation}
  \f{\rmD}{\rmD t}=\f{\p}{\p t}+n_0\f{\p}{\p\lambda_0}.
\end{equation}

Even a simplified model of collective effects has certain minimal
requirements.  The terms added to the dynamical equations should
regularize the resonant singularity that occurs in the first-order
linearized response.  At the same time they should have the correct
form to conserve angular momentum while dissipating energy.  Since
viscosity is known to be required to resolve the singularity in the
linearized equations at a corotation resonance, we add diffusive terms
to each equation for the purposes of regularization.  The diffusion
coefficient $\nu$ can be identified as the effective viscosity of the
disc.  Specifically, we adopt the model
\begin{equation}
  \f{\rmD\lambda}{\rmD t}=\f{\p H}{\p\Lambda}+
  \left(\f{\rmd m}{\rmd a_0}\right)^{-1}\f{\p}{\p a_0}
  \left(\nu\f{\rmd m}{\rmd a_0}\f{\p\lambda}{\p a_0}\right),
\end{equation}
\begin{equation}
  \f{\rmD\Lambda}{\rmD t}=-\f{\p H}{\p\lambda}+
  \left(\f{\rmd m}{\rmd a_0}\right)^{-1}\f{\p}{\p a_0}
  \left(\nu\f{\rmd m}{\rmd a_0}\f{\p\Lambda}{\p a_0}\right)-
  2\nu\left|\f{\p z_\beta}{\p a_0}\right|^2,
\label{energy}
\end{equation}
\begin{equation}
  \f{\rmD z_\alpha}{\rmD t}=-\rmi\f{\p H}{\p z_\alpha^*}+
  \left(\f{\rmd m}{\rmd a_0}\right)^{-1}\f{\p}{\p a_0}
  \left(\nu\f{\rmd m}{\rmd a_0}\f{\p z_\alpha}{\p a_0}\right),
\end{equation}
where $m(a_0)$ is the cumulative mass function of the disc, i.e.~the
total mass interior to the orbit labelled by $a_0$.  (For a circular
disc, $\rmd m/\rmd a_0=2\pi\Sigma a_0$, where $\Sigma$ is the surface
density.)  The reason for writing the diffusive terms in this way,
rather than simply as $\nu(\p^2\Lambda/\p a_0^2)$, etc., is so that
they properly conserve the mass-weighted integrated quantities.  Now
if a certain quantity obeys a diffusion equation, the square of that
quantity is both diffused and dissipated.\footnote{Specifically, if
$\p_tu=\p_{xx}u$ and $v=|u|^2$, then $\p_tv=\p_{xx}v-2|\p_xu|^2$.}  By
allowing $z_\alpha$ to diffuse we therefore allow the specific AMD
$|z_\beta|^2$ to diffuse and dissipate.  The same dissipative term is
added to the `energy' equation~(\ref{energy}) so that the specific
angular momentum $L_z=\Lambda-|z_\beta|^2$ diffuses conservatively and
does not dissipate.

To see how this model operates, consider the behaviour in the absence
of a perturber.  With suitable boundary conditions we have, after an
integration by parts,
\begin{equation}
  \f{\rmd}{\rmd t}\int|z_\beta|^2\,\rmd m=-2\int\nu\left|\f{\p
  z_\beta}{\p a_0}\right|^2\,\rmd m,
\end{equation}
showing the dissipation of the total AMD.\footnote{A slightly more
sophisticated model would allow the dissipation to be related to
gradients of $z_\alpha$ multiplied by some function of $a_0$, e.g.~if
the gradient of $\calI$, rather than that of $z_2$, is the important
quantity in a warped disc.  However, such refinements are
inconsequential in this context.}  Energy (or $\Lambda$) is also
dissipated, but the total angular momentum perpendicular to the
reference plane is conserved:
\begin{equation}
  \f{\rmd}{\rmd t}\int\left(\Lambda-|z_\beta|^2\right)\,\rmd m=0.
\end{equation}
In these integrals the mass element is
\begin{equation}
  \rmd m=\f{1}{2\pi}\f{\rmd m}{\rmd a_0}\,\rmd a_0\,\rmd\lambda_0.
\end{equation}

We may neglect the slow effects of diffusion on the smooth unperturbed
state.  Consider now a first-order equation such as
\begin{equation}
  \f{\rmD\Lambda_1}{\rmD t}=\real\left[\rmi kR\,\rme^{\rmi\phi_0}\right]+
  \left(\f{\rmd m}{\rmd a_0}\right)^{-1}\f{\p}{\p a_0}
  \left(\nu\f{\rmd m}{\rmd a_0}\f{\p\Lambda_1}{\p a_0}\right).
\end{equation}
Unlike the ordinary differential equation~(\ref{dotL1}), this is now a
partial differential equation in variables $(a_0,\lambda_0,t)$.  Since
the equation is linear and the forcing term depends on $\lambda_0$ and
$t$ only through the exponential, the solution is of the form
$\Lambda_1=\real\left[\tilde\Lambda_1(a_0)\,\rme^{\rmi\phi_0}\right]$,
where $\tilde\Lambda_1$ satisfies the ordinary differential equation
\begin{equation}
  \rmi\dot\phi_0\tilde\Lambda_1=\rmi kR+\left(\f{\rmd m}{\rmd a_0}\right)^{-1}
  \f{\rmd}{\rmd a_0}\left(\nu\f{\rmd m}{\rmd a_0}
  \f{\rmd\tilde\Lambda_1}{\rmd a_0}\right).
\end{equation}
It is clear that the viscous term is required to regularize the
solution at the resonance where $\dot\phi_0=0$.  For a localized
response the variation of all coefficients with $a_0$ can be
neglected, except for $\dot\phi_0\approx Dx$, where $x=a_0-\hat a_0$
is the distance from the resonance and $D=\rmd\dot\phi_0/\rmd a_0$ is
the rate of detuning.  Then $\tilde\Lambda_1=fkR$, where $f(x)$ now
denotes the solution (decaying as $x\to\pm\infty$) of the rescaled
equation with unit forcing,
\begin{equation}
  Dx f+\rmi\nu\f{\rmd^2f}{\rmd x^2}=1.
\end{equation}
Standard Fourier-transform methods \citep[cf.][]{MS87} give
\begin{equation}
  f(x)=\f{\rmi}{|D|}\int_{-\infty}^\infty\rme^{\rmi kx+(\nu/3D)k^3}
  H(-k\,\sgn D)\,\rmd k,
\label{fx}
\end{equation}
where $H$ is the Heaviside step function.  Note that
$f\approx(Dx)^{-1}\approx\dot\phi_0^{-1}$ far from the resonance.  In
fact $f$ resembles $(Dx-\rmi s)^{-1}$ for all $x$ (Fig.~1) if
$s\approx(\nu D^2)^{1/3}$.  Furthermore
\begin{equation}
  \int_{-\infty}^\infty f_\rmi\,\rmd x=\int_{-\infty}^\infty
  \nu\left|\f{\rmd f}{\rmd x}\right|^2\,\rmd x=\f{\pi}{|D|}.
\label{intfi}
\end{equation}
Therefore the resolution of the resonant singularity by viscosity
gives a solution with properties very similar to that of the simple
replacement~(\ref{replace}).  In particular, the integrated effect of
the imaginary part of $f$, which gives rise to the irreversible
behaviour, is independent of $\nu$ and identical to that obtained
using equation~(\ref{replace}).

\begin{figure*}
  \centerline{\epsfysize8cm\epsfbox{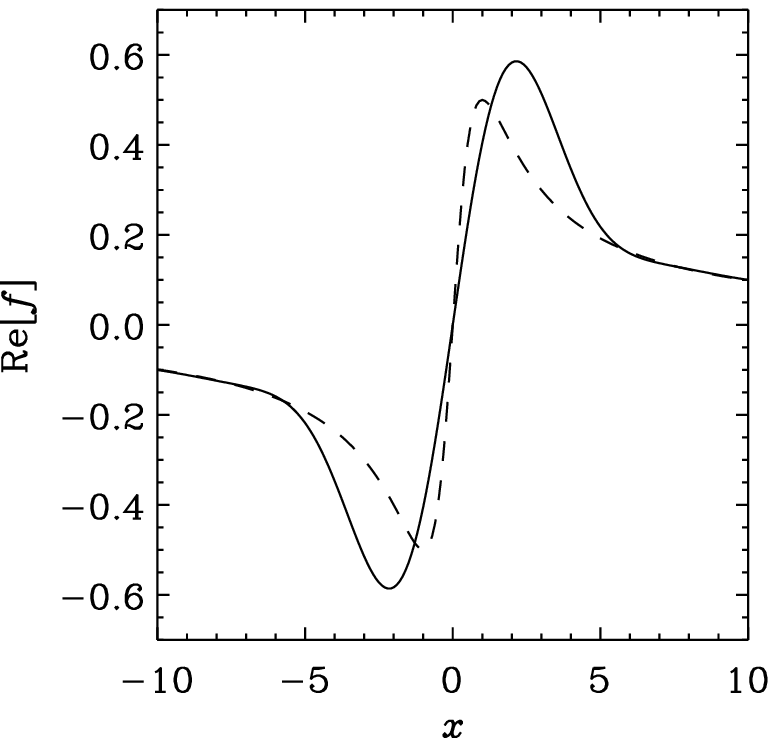}\qquad
    \epsfysize8cm\epsfbox{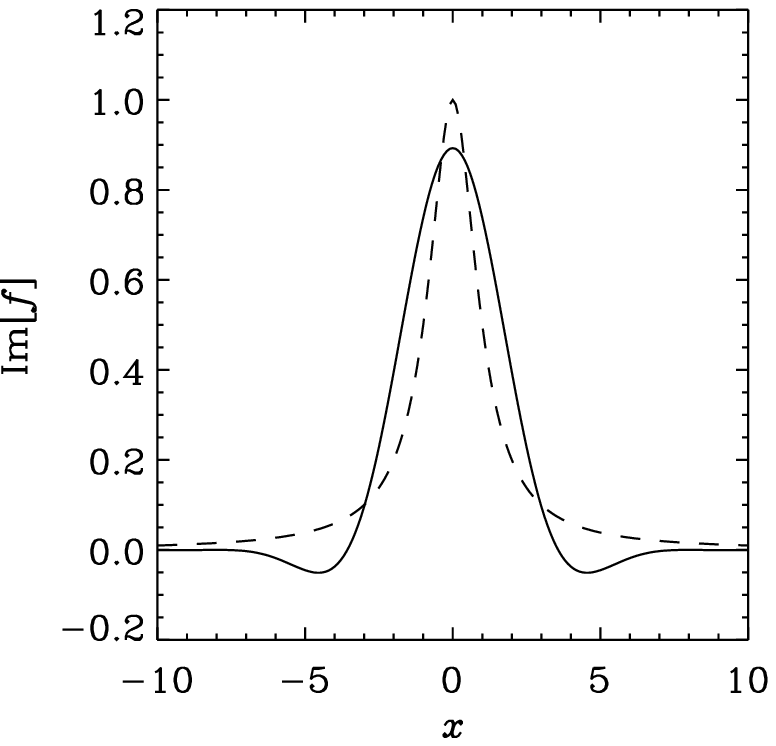}}
  \caption{Real and imaginary parts of the complex function $f(x)$,
    which represents how the resonant singularity at $x=0$ is
    resolved.  The solid line shows the viscous solution,
    equation~(\ref{fx}), while the dashed line gives the simpler
    `damped' model $f=(Dx-\rmi s)^{-1}$ for $s=1$.  Without loss of
    generality, units are chosen such that $D=\nu=1$.}
\end{figure*}

The solution for $\lambda_1$ and $z_{\alpha1}$ follows in the same
way, with the complex $f(x)$ replacing the $\dot\phi_0^{-1}$ of the
original analysis of Section~\ref{s:perturbative}.\footnote{It is easy
to generalize this treatment of collective effects to allow different
viscosities to act on the mean-motion variables $(\Lambda,\lambda)$,
the eccentricity variable $z_1$ and the inclination variable $z_2$.
Complex viscosities are also permissible, giving solutions in the form
of attenuated waves emitted from the resonance.}  The solution
proceeds similarly to the second order, with one important exception.
The dissipative term in the `energy' equation~(\ref{energy}) provides
an additional contribution to $\langle\dot\Lambda\rangle$, which is
needed to enforce angular momentum conservation.  Using
equation~(\ref{intfi}) and carrying out an integration by parts, we
find
\begin{equation}
  \int\langle\dot\Lambda\rangle_\rmd\,\rmd m=
  -\f{\pi}{2|D|}\left[(k+1)\left|\f{\p R}{\p z_\beta}\right|^2
  \f{\rmd m}{\rmd a_0}-(k-1)\left|\f{\p R}{\p z_\beta^*}\right|^2
  \f{\rmd m}{\rmd a_0}+k^2|R|^2\f{\rmd}{\rmd a_0}
  \left(\f{\rmd m}{\rmd\Lambda_0}\right)\right].
\label{torque}
\end{equation}
Here $\langle\dot\Lambda\rangle_\rmd$ includes the dissipative
contribution, which has the effect of changing $k$ into $k\pm1$.  If
the disc remains circular, $\Lambda$ is the specific angular momentum
and the above rate of change of action corresponds to the torque on
the disc.  It agrees in form with the torque formulae for Lindblad and
corotation resonances \citep{GT79}.  Note that $\rmd m/\rmd\Lambda_0$
(${}=4\pi\Sigma/n$) is inversely proportional to the vortensity of a
circular Keplerian disc, so the corotation torque is proportional to
the gradient of inverse vortensity while the Lindblad/vertical torques
are proportional to the surface density.

It appears that this procedure cannot be followed if action--angle
eccentricity and inclination variables are used as in \citet{GT81} and
\citet{BGT84}.  We have found it essential to allow eccentricity and
inclination to diffuse (or propagate) at first order, to resolve the
resonant singularity, and for the accompanying dissipation to appear
at second order.  This is most naturally achieved by using complex
variables for eccentricity and inclination, to which the energy and
AMD are related quadratically.

\section{The unrestricted problem}
\label{s:unrestricted}

\subsection{Canonical equations}

We now extend the problem to treat dynamically the orbit of the
companion of mass $M'$ around the central mass $M$.  We adopt
canonical variables $(\lambda',\Lambda',q_\alpha',p_\alpha')$ for the
two-body problem, where $\Lambda'=[G(M+M')a']^{1/2}$ and $q_\alpha'$
and $p_\alpha'$ are defined by analogy with equations
(\ref{q1})--(\ref{p2}).

The Hamiltonian of the unperturbed two-body problem is $\mu H_0'$,
where $\mu=MM'/(M+M')$ is the reduced mass and
\begin{equation}
  H_0'=-\f{[G(M+M')]^2}{2\Lambda'^2}=-\f{G(M+M')}{2a'},
\end{equation}
giving rise to an unperturbed mean motion
\begin{equation}
  \dot\lambda'=n'(\Lambda')=\f{\rmd H_0'}{\rmd\Lambda'}=
  \left[\f{G(M+M')}{a'^3}\right]^{1/2}.
\end{equation}

The total Hamiltonian of the coupled system is then of the form
\begin{equation}
  H=\int\left\{H_0(\Lambda)-\epsilon\,
  \real\left[R(\Lambda,q_\alpha,p_\alpha,\Lambda',q_\alpha',p_\alpha')\,
  \rme^{\rmi\phi}\right]\right\}\,\rmd m+\mu H_0'(\Lambda'),
\end{equation}
and Hamilton's equations read
\begin{equation}
  \dot\lambda=n(\Lambda)-\epsilon\,\real\left[\f{\p R}{\p\Lambda}\,
  \rme^{\rmi\phi}\right],
\end{equation}
\begin{equation}
  \dot\Lambda=\epsilon\,\real\left[\rmi k R\,\rme^{\rmi\phi}\right],
\end{equation}
\begin{equation}
  \dot q_\alpha=-\epsilon\,\real\left[\f{\p R}{\p p_\alpha}\,
  \rme^{\rmi\phi}\right],
\end{equation}
\begin{equation}
  \dot p_\alpha=\epsilon\,\real\left[\f{\p R}{\p q_\alpha}\,
  \rme^{\rmi\phi}\right],
\end{equation}
as for the restricted problem, together with
\begin{equation}
  \dot\lambda'=n'(\Lambda')-\f{\epsilon}{\mu}\int\real
  \left[\f{\p R}{\p\Lambda'}\,\rme^{\rmi\phi}\right]\,\rmd m,
\end{equation}
\begin{equation}
  \dot\Lambda'=\f{\epsilon}{\mu}\int\real
  \left[\rmi k'R\,\rme^{\rmi\phi}\right]\,\rmd m,
\end{equation}
\begin{equation}
  \dot q_\alpha'=-\f{\epsilon}{\mu}\int\real
  \left[\f{\p R}{\p p_\alpha'}\,\rme^{\rmi\phi}\right]\,\rmd m,
\end{equation}
\begin{equation}
  \dot p_\alpha'=\f{\epsilon}{\mu}\int\real
  \left[\f{\p R}{\p q_\alpha'}\,\rme^{\rmi\phi}\right]\,\rmd m,
\end{equation}
determining the feedback on the two-body orbit.

\subsection{Perturbative solution}

The solution can be developed in powers of $\epsilon$ as in
Section~\ref{s:perturbative}.  When it comes to calculating the
secular evolution at $O(\epsilon^2)$ there are no cross-contributions
from different regions of the disc because their first-order
oscillations are out of phase.  We then find that the secular dynamics
at second order, away from resonance, is governed by the Hamiltonian
\begin{equation}
  \int\calH\,\rmd m
\end{equation}
where $\calH$ is given by equation~(\ref{calH}) or
equation~(\ref{hz}), as in the restricted problem, except that $\calH$
is now a function of
$(\Lambda,q_\alpha,p_\alpha,\Lambda',q_\alpha',p_\alpha')$.  In
detail,
\begin{equation}
  \langle\dot\Lambda\rangle=\f{\p\calH}{\p\lambda}=0,
\end{equation}
\begin{equation}
  \rmi\langle\dot z_\alpha\rangle=\f{\p\calH}{\p z_\alpha^*},
\end{equation}
\begin{equation}
  \langle\dot\Lambda'\rangle=\f{1}{\mu}\int\f{\p\calH}{\p\lambda'}\,\rmd m=0,
\end{equation}
\begin{equation}
  \rmi\langle\dot z_\alpha'\rangle=\f{1}{\mu}\int\f{\p\calH}{\p z_\alpha^*}\,
  \rmd m.
\end{equation}

The total AMD of the system is conserved under
this precessional dynamics because
\begin{equation}
  \bigg\langle\f{\rmd}{\rmd t}\left(\int|z_\beta|^2\,\rmd m+
  \mu|z_\beta'|^2\right)\bigg\rangle=\rmi\int\left(z_\beta
  \f{\p\calH}{\p z_\beta}-z_\beta^*\f{\p\calH}{\p z_\beta^*}+z_\beta'
  \f{\p\calH}{\p z_\beta'}-z_\beta'^*\f{\p\calH}{\p z_\beta'^*}\right)\,
  \rmd m=-\int\f{\p\calH}{\p\theta}\,\rmd m=0,
\end{equation}
where $\theta$ is an angle of rotation of the entire system about the
$z$-axis.  (Under a rotation through $\delta\theta$, $\varpi$ and
$\Omega$ decrease by $\delta\theta$ and so $\delta z_\alpha=-\rmi
z_\alpha\,\delta\theta$.)  This conservation law is in addition to the
conservation of $\int\calH\,\rmd m$ itself.

Resolution of the resonant singularity can again be achieved using the
replacement (\ref{replace}) or a more sophisticated version thereof.
We find
\begin{equation}
  \langle\dot\Lambda\rangle=-\f{1}{2}kf_\rmi\left(\left|\f{\p R}{\p z_\beta}
  \right|^2-\left|\f{\p R}{\p z_\beta^*}\right|^2\right)+\f{1}{2}k^2
  \f{\p}{\p\Lambda}(f_\rmi|R|^2),
\label{lambdadot}
\end{equation}
\begin{equation}
  \rmi\langle\dot z_\alpha\rangle=\f{1}{4}f\left(\f{\p^2R^*}
  {\p z_\alpha^*\p z_\beta^*}\f{\p R}{\p z_\beta}-
  \f{\p^2R^*}{\p z_\alpha^*\p z_\beta}\f{\p R}{\p z_\beta^*}\right)+
  \f{1}{4}f^*\left(\f{\p^2R}{\p z_\alpha^*\p z_\beta}\f{\p R^*}{\p z_\beta^*}-
  \f{\p^2R}{\p z_\alpha^*\p z_\beta^*}\f{\p R^*}{\p z_\beta}\right)-
  \f{1}{4}k\f{\p}{\p\Lambda}\left(fR\f{\p R^*}{\p z_\alpha^*}+
  f^*R^*\f{\p R}{\p z_\alpha^*}\right),
\end{equation}
\begin{equation}
  \langle\dot\Lambda'\rangle=\f{1}{\mu}\int\bigg[-\f{1}{2}k'f_\rmi
  \left(\left|\f{\p R}{\p z_\beta}\right|^2-
  \left|\f{\p R}{\p z_\beta^*}\right|^2\right)+
  \f{1}{2}kk'\f{\p}{\p\Lambda}(f_\rmi|R|^2)\bigg]\,\rmd m,
\label{lambdaprimedot}
\end{equation}
\begin{eqnarray}
  \lefteqn{\rmi\langle\dot z_\alpha'\rangle=\f{1}{\mu}\int\bigg[\f{1}{4}f
  \left(\f{\p^2R^*}{\p z_\alpha'^*\p z_\beta^*}\f{\p R}{\p z_\beta}-
  \f{\p^2R^*}{\p z_\alpha'^*\p z_\beta}\f{\p R}{\p z_\beta^*}\right)+
  \f{1}{4}f^*\left(\f{\p^2R}{\p z_\alpha'^*\p z_\beta}\f{\p R^*}{\p z_\beta^*}-
  \f{\p^2R}{\p z_\alpha'^*\p z_\beta^*}\f{\p R^*}{\p z_\beta}\right)}&
  \nonumber\\
  &&\qquad-\f{1}{4}k\f{\p}{\p\Lambda}\left(fR\f{\p R^*}{\p z_\alpha'^*}+
  f^*R^*\f{\p R}{\p z_\alpha'^*}\right)\bigg]\,\rmd m.
\end{eqnarray}
Equation~(\ref{lambdadot}) requires modification to account for
dissipation, as discussed in Section~\ref{s:collective}, leading again
to equation~(\ref{torque}).

According to this analysis the secular behaviour of the disc is
governed by the precisely same equations as in the restricted problem,
but the orbital elements of the companion now also evolve in a
consistent way through the feedback of the disc on the two-body orbit.
This coupled dynamics is of special interest when the disc and
companion have comparable angular momenta, as may occur in
protoplanetary systems or planetary rings.

\section{Applications}
\label{s:applications}

We now work out some of the implications of the preceding analysis.
We restrict attention initially to effects of first degree in
eccentricity and inclination, and then consider an example of nonlinear
behaviour in Section~\ref{s:beyond}.

\subsection{Second-order resonances}

In Section~\ref{s:resolution} we considered the behaviour of
eccentricity at an interior second-order resonance in the circular
restricted problem.  We now generalize the result to permit both the
disc and the companion to have small eccentricities and inclinations,
and to allow dynamical feedback on the companion's orbit.

For an interior second-order $j:j-2$ resonance we have, to lowest
degree in eccentricity and inclination,
\begin{equation}
  \rmi\langle\dot z_1\rangle=\f{1}{4}\left(\f{GM'}{a'}\right)^2
  \left(\f{2}{\Lambda_0}\right)^{3/2}f
  \left[2f_{45}(2f_{45}\calE+f_{49}\calE')\right],
\end{equation}
\begin{equation}
  \rmi\langle\dot z_2\rangle=\f{1}{4}\left(\f{GM'}{a'}\right)^2
  \left(\f{2}{\Lambda_0}\right)^{3/2}f
  \left[\f{1}{4}f_{57}^2(\calI-\calI')\right],
\end{equation}
\begin{equation}
  \rmi\langle\dot z_1'\rangle=\f{1}{\mu}\int\f{1}{4}\left(\f{GM'}{a'}\right)^2
  \left(\f{2}{\Lambda_0'}\right)^{1/2}\left(\f{2}{\Lambda_0}\right)f
  \left[f_{49}(2f_{45}\calE+f_{49}\calE')\right]\,\rmd m,
\end{equation}
\begin{equation}
  \rmi\langle\dot z_2'\rangle=\f{1}{\mu}\int\f{1}{4}\left(\f{GM'}{a'}\right)^2
  \left(\f{2}{\Lambda_0'}\right)^{1/2}\left(\f{2}{\Lambda_0}\right)f
  \left[\f{1}{4}f_{57}^2(\calI'-\calI)\right]\,\rmd m.
\end{equation}
These expressions, which follow from the terms of second degree in the
disturbing function~(\ref{second-interior}), derive only from the
`Lindblad/vertical' aspects of the resonance; there are no
`corotation' terms here.  The rate of change of total AMD is
\begin{equation}
  \bigg\langle\f{\rmd}{\rmd t}
  \left(\int|z_\beta|^2\,\rmd m+\mu|z_\beta'|^2\right)\bigg\rangle=
  \int\f{1}{4}\left(\f{GM'}{a'}\right)^2
  \left(\f{2}{\Lambda_0}\right)2f_\rmi
  \Big(|2f_{45}\calE+f_{49}\calE'|^2+\f{1}{4}f_{57}^2|\calI-\calI'|^2\Big)\,
  \rmd m.
\end{equation}
For a localized resonance $f_\rmi$ is strongly peaked and we may use
equation~(\ref{intfi}) to express the result as
\begin{equation}
  \left(\f{GM'}{a'}\right)^2\f{\pi}{\Lambda_0|D|}\f{\rmd m}{\rmd a_0}
  \left[|2f_{45}\calE+f_{49}\calE'|^2+\f{1}{4}f_{57}^2|\calI-\calI'|^2\right],
\label{elr}
\end{equation}
evaluated at the resonance.  The equivalent result for an exterior
second-order resonance is
\begin{equation}
  \left(\f{GM'}{a}\right)^2\f{\pi}{\Lambda_0|D|}\f{\rmd m}{\rmd a_0}
  \left[|2(f_{53}-{\textstyle\f{3}{8}}\alpha^{-2}\delta_{j,3})\calE+
  f_{49}\calE'|^2+\f{1}{4}f_{57}^2|\calI-\calI'|^2\right].
\end{equation}

The meaning of these results is that second-order resonances allow an
exponential growth of AMD for small eccentricities and inclinations.
We have already explained the connection with the result of \citet{L91}
concerning the growth of eccentricity in the disc when the companion
has a fixed circular orbit.  (It is noteworthy that the indirect term
affects, and largely cancels, the eccentricity growth rate associated
with the $1:3$ resonance in an exterior disc, so it should always be
included in studies of satellite--disc interactions.)  On the other
hand, if the disc is fixed artificially to be circular, the
eccentricity of the companion can grow exponentially.  Using the fact
that $f_{49}\approx-(j^2/2\pi)[5K_0(4/3)+(19/4)K_1(4/3)]$ for $j\gg1$,
where $K$ is the modified Bessel function, we find a growth rate for
$\calE'$ that agrees with equation~(8) of the analysis of eccentric
Lindblad resonances by \citet{W88}\footnote{In his equation the second
$K_1$ should be $K_0$.} in the case of a circular disc and $e'\ll1$.

For the first time, though, we see here the effect of the resonance in
the general case when both the disc and companion are eccentric.
Certain linear combinations of the complex variables $\calE$ and
$\calE'$ are involved, which depend on $j$.  The case of inclination
is easier to understand, as the relevant combination is always
$\calI-\calI'$, the complex `mutual inclination' of the disc and
companion.  Indeed, \citet{LO01} were able to write down such
equations for inclination dynamics based on the formulae of
\citet{BGT84} (for a fixed disc) and the idea that only the complex
mutual inclination could enter.  Using the fact that
$f_{57}=(\alpha/2)b_{3/2}^{(j-1)}(\alpha)$, we find agreement with
equation~(48) of \citet{LO01}.  Although $2f_{45}/f_{49}\ne-1$, it
approaches $-1$ for large $j$, and so a similar combination
$\calE-\calE'$ is involved in that limit.  Especially in a
protoplanetary system in which the disc and planet(s) have comparable
orbital angular momenta and engage in coupled secular dynamics, the
eccentricities and inclinations of both disc and companion(s) must be
considered simultaneously.

\subsection{First-order resonances}

In the case of a circular and coplanar system the only operative term
in the disturbing function for an interior first-order $j:j-1$
resonance is
\begin{equation}
  R=\f{GM'}{a'}f_{27}\calE.
\end{equation}
This leads to no evolution of $z_\alpha$ or $z_\alpha'$ (which remain
zero) but does produce a torque on the disc, which is (according to
equation~\ref{torque})
\begin{equation}
  -\f{\pi j}{\Lambda_0|D|}\left(\f{GM'}{a'}\right)^2f_{27}^2
  \f{\rmd m}{\rmd a_0}.
\end{equation}
Using the fact that $f_{27}=-jb_{1/2}^{(j)}(\alpha)-(\alpha/2)\rmd
b_{1/2}^{(j)}/\rmd\alpha$, this torque can be shown to agree with
equation~(13) of \citet{GT80}.  The equal and opposite torque on the
companion follows from equation~(\ref{lambdaprimedot}).

When small eccentricities and inclinations are allowed for we obtain
the following dynamics to lowest degree:
\begin{eqnarray}
  \lefteqn{\rmi\langle\dot z_1\rangle=\f{1}{4}\left(\f{GM'}{a'}\right)^2
  \left(\f{2}{\Lambda_0}\right)^{3/2}f_{27}\left\{
  f\left[2(f_{28}-{\textstyle\f{1}{8}}f_{27})\calE+2f_{35}\calE'\right]+
  f^*\left[2(f_{28}-{\textstyle\f{1}{8}}f_{27})\calE+
  (f_{32}+\alpha\delta_{j,2})\calE'\right]\right\}}&\nonumber\\
  &&-\f{1}{4}(j-1)\left(\f{GM'}{a'}\right)^2\left\{\f{\p}{\p\Lambda_0}
  \left[\left(\f{2}{\Lambda_0}\right)ff_{27}^2\right]z_1+\f{\p}{\p\Lambda_0}
  \left[\left(\f{2}{\Lambda_0}\right)^{1/2}\left(\f{2}{\Lambda_0'}\right)^{1/2}
  ff_{27}(f_{31}-2\alpha\delta_{j,2})\right]z_1'\right\},
\end{eqnarray}
\begin{equation}
  \rmi\langle\dot z_2\rangle=\f{1}{4}\left(\f{GM'}{a'}\right)^2
  \left(\f{2}{\Lambda_0}\right)^{3/2}f_{27}\left\{
  f\left[-{\textstyle\f{1}{8}}f_{39}\calI+
  {\textstyle\f{1}{8}}f_{40}(2\calI'-\calI)\right]+
  f^*\left[{\textstyle\f{1}{8}}f_{39}(2\calI'-\calI)-
  {\textstyle\f{1}{8}}f_{40}\calI\right]\right\},
\end{equation}
\begin{eqnarray}
  \lefteqn{\rmi\langle\dot z_1'\rangle=\f{1}{\mu}\int\f{1}{4}
  \left(\f{GM'}{a'}\right)^2
  \left(\f{2}{\Lambda_0'}\right)^{1/2}\left(\f{2}{\Lambda_0}\right)f_{27}
  \left\{f\left[f_{29}\calE'+(f_{32}+\alpha\delta_{j,2})\calE\right]+
  f^*\left[f_{29}\calE'+2f_{35}\calE\right]\right\}\,\rmd m}&\nonumber\\
  &&-\f{1}{\mu}\int\f{1}{4}(j-1)\left(\f{GM'}{a'}\right)^2\left\{
  \f{\p}{\p\Lambda_0}
  \left[\left(\f{2}{\Lambda_0}\right)^{1/2}\left(\f{2}{\Lambda_0'}\right)^{1/2}
  ff_{27}(f_{31}-2\alpha\delta_{j,2})\right]z_1+\f{\p}{\p\Lambda_0}
  \left[\left(\f{2}{\Lambda_0'}\right)f(f_{31}-2\alpha\delta_{j,2})^2\right]
  z_1'\right\}\,\rmd m,\nonumber\\
\end{eqnarray}
\begin{equation}
  \rmi\langle\dot z_2'\rangle=\f{1}{\mu}\int\f{1}{4}\left(\f{GM'}{a'}\right)^2
  \left(\f{2}{\Lambda_0'}\right)^{1/2}\left(\f{2}{\Lambda_0}\right)f_{27}
  \left\{f\left[{\textstyle\f{1}{8}}f_{39}(2\calI-\calI')-
  {\textstyle\f{1}{8}}f_{40}\calI'\right]+
  f^*\left[-{\textstyle\f{1}{8}}f_{39}\calI'+
  {\textstyle\f{1}{8}}f_{40}(2\calI-\calI')\right]\right\}\,\rmd m.
\end{equation}
Both `Lindblad' and `corotation' aspects of the resonance enter here.
Regarding the `corotation' terms (those involving derivatives with
respect to $\Lambda_0$), we again have precessional behaviour that
diverges, this time more strongly, at the resonance, owing to the
appearance of $\rmd f/\rmd\Lambda_0$.  Resolution of the resonant
singularity leads to an eccentricity growth rate (e.g.~when
$\calE'=0$) that changes sign across the location of the resonance.
Which sign is obtained for the disc as a whole depends on whether
$\rmd m/\rmd\Lambda_0$ is greater on one side or the other.  The net
effect for a narrow resonance is proportional to
$\rmd^2m/\rmd\Lambda_0^2$, or equivalently $\rmd(\Sigma/n)/\rmd a_0$,
as is characteristic of corotation resonances.

It is notable that $\langle\dot z_2\rangle$ and $\langle\dot
z_2'\rangle$ do not vanish when the disc and companion have inclined
but coplanar orbits ($\calI=\calI'\ne0$).  Using the relation
$f_{39}-f_{40}=2jf_{27}$ it is possible to show that $\langle\dot
z_2\rangle$ and $\langle\dot z_2'\rangle$ in this case can be
attributed entirely to $\langle\dot\Lambda\rangle$ and
$\langle\dot\Lambda'\rangle$ while $\calI$ and $\calI'$ remain
constant.  In general, the rate of change of total AMD is
\begin{equation}
  \left(\f{GM'}{a'}\right)^2\f{(j-1)\pi}{2|D|}\f{\rmd}{\rmd a_0}
  \left(\f{\rmd m}{\rmd\Lambda_0}\right)
  |f_{27}\calE+(f_{31}-2\alpha\delta_{j,2})\calE'|^2.
\label{ecr}
\end{equation}
Note that only the `corotation' terms contribute to this expression.
The equivalent result for an exterior first-order resonance is
\begin{equation}
  -\left(\f{GM'}{a}\right)^2\f{j\pi}{2|D|}\f{\rmd}{\rmd a_0}
  \left(\f{\rmd m}{\rmd\Lambda_0}\right)
  |f_{27}\calE'+(f_{31}-{\textstyle\f{1}{2}}\alpha^{-2}\delta_{j,2})\calE|^2.
\end{equation}
Using the fact that
$f_{31}\approx-f_{27}\approx(j/\pi)[2K_0(2/3)+K_1(2/3)]$ for $j\gg1$,
we find a growth (or decay) rate for $\calE'$ that agrees with
equation~(12) of the analysis of eccentric corotation resonances by
\citet{W88} in the case of a circular disc and $e'\ll1$.  Whether the
AMD grows or decays depends on the sign of the vortensity gradient.
In general, as for second-order resonances, the eccentricities of both
disc and companion must be considered simultaneously, as the resonant
effect depends on a linear combination of $\calE$ and $\calE'$, in
fact $\calE-\calE'$ in the limit of large $j$.

\subsection{Zeroth-order resonances}

Zeroth-order resonances are coorbital and are active only if a clean
gap is not opened by the companion's orbit.  In the case of a circular
and coplanar system the only operative term in the disturbing function
for a zeroth-order resonance is
\begin{equation}
  R=\f{GM'}{a'}(2f_1-\alpha\delta_{j,1}).
\end{equation}
This leads to no evolution of $z_\alpha$ or $z_\alpha'$ (which remain
zero) but does produce a torque on the disc, which is (according to
equation~\ref{torque})
\begin{equation}
  -\f{\pi j^2}{2|D|}\left(\f{GM'}{a'}\right)^2(2f_1-\alpha\delta_{j,1})^2
  \f{\rmd}{\rmd a_0}\left(\f{\rmd m}{\rmd\Lambda_0}\right).
\end{equation}
Using the fact that $f_1={\textstyle\f{1}{2}}b_{1/2}^{(j)}(\alpha)$,
this torque can be shown to agree with equation~(14) of \citet{GT80},
and the equal and opposite torque on the companion follows from
equation~(\ref{lambdaprimedot}).  Setting $\alpha=1$ renders the
coefficients singular, and this must be resolved in practice by a
smoothing process that represents an averaging of the potential over
the vertical extent of the disc, which we do not consider here.

When small eccentricities and inclinations are allowed for we obtain
the following dynamics to lowest degree:
\begin{equation}
  \rmi\langle\dot z_1\rangle=\f{1}{4}\left(\f{GM'}{a'}\right)^2
  \left(\f{2}{\Lambda_0}\right)^{3/2}
  (2f_2+{\textstyle\f{1}{2}}\alpha\delta_{j,1})\left\{
  -f\left[(2f_2+{\textstyle\f{1}{2}}\alpha\delta_{j,1})\calE+
  (\tilde f_{10}-\alpha\delta_{j,2})\calE'\right]+
  f^*\left[(2f_2+{\textstyle\f{1}{2}}\alpha\delta_{j,1})\calE+
  f_{10}\calE'\right]\right\},
\end{equation}
\begin{eqnarray}
  \lefteqn{\rmi\langle\dot z_2\rangle=\f{1}{4}\left(\f{GM'}{a'}\right)^2
  \left(\f{2}{\Lambda_0}\right)^{3/2}{\textstyle\f{1}{4}}
  (2f_3+\alpha\delta_{j,1})\left\{
  -f\left[{\textstyle\f{1}{4}}(2f_3+\alpha\delta_{j,1})\calI+
  {\textstyle\f{1}{4}}(\tilde f_{14}-2\alpha\delta_{j,1})\calI'\right]+
  f^*\left[{\textstyle\f{1}{4}}(2f_3+\alpha\delta_{j,1})\calI+
  {\textstyle\f{1}{4}}f_{14}\calI'\right]\right\},}&\nonumber\\
\end{eqnarray}
\begin{eqnarray}
  \lefteqn{\rmi\langle\dot z_1'\rangle=\f{1}{\mu}\int\f{1}{4}
  \left(\f{GM'}{a'}\right)^2
  \left(\f{2}{\Lambda_0'}\right)^{1/2}\left(\f{2}{\Lambda_0}\right)
  \big\{-f(\tilde f_{10}-\alpha\delta_{j,2})\left[
  (2f_2+{\textstyle\f{1}{2}}\alpha\delta_{j,1})\calE+
  (\tilde f_{10}-\alpha\delta_{j,2})\calE'\right]}&\nonumber\\
  &&\qquad+f^*f_{10}\left[(2f_2+{\textstyle\f{1}{2}}\alpha\delta_{j,1})\calE+
  f_{10}\calE'\right]\big\}\,\rmd m,
\end{eqnarray}
\begin{eqnarray}
  \lefteqn{\rmi\langle\dot z_2'\rangle=\f{1}{\mu}\int\f{1}{4}
  \left(\f{GM'}{a'}\right)^2
  \left(\f{2}{\Lambda_0'}\right)^{1/2}\left(\f{2}{\Lambda_0}\right)
  \big\{-f{\textstyle\f{1}{4}}(2\tilde f_{14}-2\alpha\delta_{j,1})
  \left[{\textstyle\f{1}{4}}(2f_3+\alpha\delta_{j,1})\calI+
  {\textstyle\f{1}{4}}(\tilde f_{14}-2\alpha\delta_{j,1})\calI'\right]}&
  \nonumber\\
  &&\qquad+f^*{\textstyle\f{1}{4}}f_{14}\left[{\textstyle\f{1}{4}}
  (2f_3+\alpha\delta_{j,1})\calI+{\textstyle\f{1}{4}}f_{14}\calI'\right]\big\}
  \,\rmd m.
\end{eqnarray}
For brevity we omit `corotation' effects here and focus on the
`Lindblad' terms.  The rate of change of total AMD is
\begin{eqnarray}
  \lefteqn{-\left(\f{GM'}{a'}\right)^2\f{\pi}{\Lambda_0|D|}\f{\rmd m}{\rmd a_0}
  \Big[|(2f_2+{\textstyle\f{1}{2}}\alpha\delta_{j,1})\calE+
  (\tilde f_{10}-\alpha\delta_{j,2})\calE'|^2+
  |(2f_2+{\textstyle\f{1}{2}}\alpha\delta_{j,1})\calE+f_{10}\calE'|^2}&
  \nonumber\\
  &&\qquad+|{\textstyle\f{1}{4}}(2f_3+\alpha\delta_{j,1})\calI+
  {\textstyle\f{1}{4}}(\tilde f_{14}-2\alpha\delta_{j,1})\calI'|^2+
  |{\textstyle\f{1}{4}}(2f_3+\alpha\delta_{j,1})\calI+
  {\textstyle\f{1}{4}}f_{14}\calI'|^2\Big],
\end{eqnarray}
again omitting `corotation' terms.  Although these expressions cannot
be consistently evaluated without a softening procedure and the
further considerations of \citet{W88}, we see that the effect of the
coorbital `Lindblad' terms is to damp the eccentricity and inclination
of the disc or companion if the other is fixed at zero eccentricity
and inclination.  Generally, however, the effect of the resonance
depends on linear combinations of $\calE$ and $\calE'$, and of $\calI$
and $\calI'$, which are again $\calE-\calE'$ and $\calI-\calI'$ in the
limit of large $j$.  In this sense it is the (complex) differences in
eccentricities and inclinations of the disc and companion that are
being damped, so these terms attempt to equalize both $e$ and $\varpi$
(or $i$ and $\Omega$) of the disc and companion.

Again, it is notable that $\langle\dot z_2\rangle$ and $\langle\dot
z_2'\rangle$ do not vanish when the disc and companion have inclined
but coplanar orbits ($\calI=\calI'\ne0$).  Using the relations
$2f_3+f_{14}+\alpha\delta_{j,1}=-2f_3-\tilde
f_{14}+\alpha\delta_{j,1}=-j(2f_1-\alpha\delta_{j,1})$ it is possible
to show that $\langle\dot z_2\rangle$ and $\langle\dot z_2'\rangle$ in
this case can be attributed entirely to $\langle\dot\Lambda\rangle$
and $\langle\dot\Lambda'\rangle$ while $\calI$ and $\calI'$ remain
constant.

\subsection{Effects beyond the first degree in eccentricity}
\label{s:beyond}

For a companion on a circular orbit (e.g.~in a close binary star) and
a coplanar disc, the disturbing function for an interior second-order
resonance is
\begin{equation}
  R=\f{GM'}{a'}\left[f_{45}\calE^2+
  (f_{46}-{\textstyle\f{1}{4}}f_{45})|\calE|^2\calE^2+O(\calE^6)\right].
\end{equation}
In this case
\begin{eqnarray}
  \lefteqn{\rmi\langle\dot z_1\rangle=\f{1}{4}\left(\f{GM'}{a'}\right)^2
  \left(\f{2}{\Lambda_0}\right)^2f_{45}\left[4ff_{45}+6(3f+f^*)
  (f_{46}-{\textstyle\f{1}{4}}f_{45})|\calE|^2\right]z_1-
  \f{1}{4}(j-2)\left(\f{GM'}{a'}\right)^2\f{\p}{\p\Lambda_0}
  \left[\left(\f{2}{\Lambda_0}\right)^22ff_{45}^2\right]|z_1|^2z_1}&\nonumber\\
  &&\quad+O(z^5).
\end{eqnarray}
The effect on the AMD of the disc is
\begin{eqnarray}
  \lefteqn{\bigg\langle\f{\rmd}{\rmd t}\int|z_\beta|^2\,\rmd m\bigg\rangle
  =\left(\f{GM'}{a'}\right)^2\f{\pi}{\Lambda_0|D|}
  \bigg\{\f{\rmd m}{\rmd a_0}f_{45}\left[4f_{45}|\calE|^2+
  12(f_{46}-{\textstyle\f{1}{4}}f_{45})|\calE|^4\right]+
  (j-2)\Lambda_0\f{\rmd}{\rmd a_0}\left(\f{\rmd m}{\rmd\Lambda_0}\right)
  f_{45}^2|\calE|^4\bigg\}}&\nonumber\\
  &&\qquad+O(z^6),
\end{eqnarray}
evaluated at the resonance.

As a specific example, for the 3:1 resonance we have $f_{45}=0.5988$
and $f_{46}=-0.1936$.  The terms in braces in the above equation are
therefore
\begin{equation}
  1.434e^2\f{\rmd m}{\rmd a_0}
  \left[1-1.470e^2-{\textstyle\f{1}{2}}\beta e^2+O(e^4)\right],
\end{equation}
where $\beta$ is the logarithmic derivative of vortensity with
respect to $a$.  This result implies that the eccentricity growth rate
is substantially reduced when $e$ is not small, providing a possible
means to limit the growth of eccentricity in superhump systems.

To evaluate consistently the evolution to higher degrees in
eccentricity and inclination, further resonances must be taken into
account.  For an $n$th-order resonance ($n\ge2$) we have
$R\propto\calE^n$, leading to a growth of eccentricity in the disc
with $\dot\calE\propto|\calE|^{2n-4}\calE$.

\section{Summary and conclusions}
\label{s:conclusions}

We have presented a new and general analysis of mean-motion resonances
between a Keplerian disc and an orbiting companion.  The emphasis of
this treatment is to provide a systematic method to calculate the
rates of change of eccentricity and inclination variables of the disc
and companion associated with resonances of various orders.

Near a mean-motion resonance, terms in the disturbing function
corresponding to the appropriate commensurability produce an
oscillation of the orbital elements of the disc and companion at first
order in the perturbation.  At second order, a secular effect is
obtained in the form of a precessional dynamics that diverges as the
resonance is approached.  When the divergence is resolved by taking
into account the (dissipative) collective effects of the disc, a
finite precession rate is obtained, together with an irreversible
evolution, such as a growth or decay of eccentricity or inclination.

Our work differs from earlier methods in which the perturbing
potential is decomposed into rigidly rotating components and the
eccentricity evolution is deduced indirectly from resonant torques
\citep{GT80,W88}.  Such methods do not apply to general
satellite--disc interactions in which both the disc and companion have
eccentric and/or inclined orbits.  Where appropriate, however, we
obtain full agreement with this earlier work.  By using the classical
disturbing function, we benefit from the ready availability of
expansions to high degrees in eccentricity and inclination.

After considerable thought, we recommend the use of complex canonical
Poincar\'e variables, as defined in Section~\ref{s:complex}, to
describe the eccentricity and inclination dynamics.  These allow the
most compact expression of the dynamical equations and the disturbing
function.  They are also directly related to the angular momentum
deficit (AMD), which is a quantity of fundamental importance, being a
positive-definite measure of the departure of the system from
circularity and coplanarity that is conserved in first-order secular
interactions.  By concentrating on the evolution of the system's AMD
we can separate the effects that lead to irreversible growth or decay
of eccentricity and inclination from those that are merely
precessional in character.  Moreover, the use of complex variables
leads, in the case of small eccentricities and inclinations, to linear
evolutionary equations that would be extremely cumbersome in any other
representation.

This work does not address certain refinements that are necessary for
a more complete description of resonant interactions.  The important
omitted effects are nonlinearity (in particular the saturation of
corotation resonances), torque cutoff effects, resonant shifts and
vertical averaging.  It is likely that all of these could be addressed
within the present framework.

The equations derived here do not stand alone but should be combined,
in future work, with the complex one-dimensional partial differential
equations describing the secular evolution of eccentricity and
inclination in the disc.  In simple cases such a programme has already
been carried out \citep{LO01,GO06}.  For small eccentricity and
inclination the outcome is a set of secular normal modes, being a
continuum analogue of the Laplace--Lagrange secular theory, which
describe rigidly precessing configurations of disc(s) and
companion(s).  The contribution of mean-motion resonances allows these
modes to grow (or decay), and any growth must be offset against the
viscous damping of the mode within the disc.  Coupled eccentric modes
have also been discussed by \cite{P02}.

Previous discussions of eccentricity growth in protoplanetary systems,
for example, have presented an interesting but incomplete picture.
\citet{GT80}, \citet{W88} and \citet{GS03} discuss the close
competition between eccentric Lindblad resonances and eccentric
corotation resonances.  This is equivalent to comparing
equations~(\ref{elr}) and~(\ref{ecr}) with the complex eccentricity
$\calE$ of the disc set to zero (and $\calI=\calI'=0)$.  Since the
disc becomes eccentric through secular (and possibly resonant)
interactions, the relative weighting of all these resonances is
adjusted in a way that depends on the eccentricity distribution within
the disc.  A self-consistent solution of the coupled secular dynamics,
as carried out by \citet{LO01} in the case of inclination, is
therefore required to address the issue.  It remains to be seen
whether protoplanetary systems support growing eccentric modes in the
presence of viscous damping \citep[see also][]{LO06}.

Another application of this work is to close binary stars, where
accretion discs can sometimes become eccentric through the action of
the 3:1 resonance.  This work suggests various elaborations of the
description by \citet{L91}.  In addition to eccentricity growth, the
3:1 resonance provides an second-order precession that may affect the
shape of the eccentric mode and alter the global precession rate of
the disc.  The growth rate is weakened when the eccentricity is no
longer small, leading to a possible saturation mechanism for the
superhump instability.  Finally, higher-order resonances such as $4:1$
provide eccentricity growth of the form
$\dot\calE\propto|\calE|^2\calE$ which, although not leading to a
linear instability, might conceivably support a global eccentric mode
in binary stars of larger mass ratio, where the disc is too small to
contain the 3:1 resonance.

\section*{acknowledgments}

I thank the referee for suggesting some improvements in the
presentation.

{}

\appendix

\section{Expansion of the disturbing function}
\label{s:df}

In this section the disturbing function is given for various
commensurabilities, correct to fourth degree in eccentricity and
inclination.  The coefficients $f_i$ are defined in Appendix~B of
\citet{MD99}.

\subsection{Secular terms}

\begin{eqnarray}
  \lefteqn{R_{0,0}=\f{GM'}{a'}\big\{f_1+f_2\left[|\calE|^2+|\calE'|^2-
  |\calI-\calI'|^2+{\textstyle\f{1}{4}}\left(|\calE|^2+|\calE'|^2\right)
  \left(\calI\calI'^*+\calI^*\calI'\right)-{\textstyle\f{1}{2}}
  \left(|\calE|^2|\calI|^2+|\calE'|^2|\calI'|^2\right)\right]}&\nonumber\\
  &&+(f_4-{\textstyle\f{1}{4}}f_2)|\calE|^4+f_5\left[|\calE|^2|\calE'|^2-
  \left(|\calE|^2+|\calE'|^2\right)|\calI-\calI'|^2\right]+
  (f_6-{\textstyle\f{1}{4}}f_2)|\calE'|^4+{\textstyle\f{1}{2}}f_{17}
  \left(\calE^2\calE'^{*2}+\calE^{*2}\calE'^2\right)\nonumber\\
  &&+{\textstyle\f{1}{2}}\left(\calE\calE'^*+\calE^*\calE'\right)
  \left[f_{10}+(f_{11}-{\textstyle\f{1}{8}}f_{10})|\calE|^2+
  (f_{12}-{\textstyle\f{1}{8}}f_{10})|\calE'|^2+{\textstyle\f{1}{4}}f_{13}
  |\calI-\calI'|^2\right]+{\textstyle\f{1}{16}}f_8|\calI-\calI'|^4
  \nonumber\\
  &&+{\textstyle\f{1}{16}}(f_8-4f_5)\left[\left(|\calI|^2+|\calI'|^2\right)
  \left(\calI\calI'^*+\calI^*\calI'\right)-4|\calI|^2|\calI'|^2\right]+
  {\textstyle\f{1}{8}}f_{18}\left[\calE^2(\calI^*-\calI'^*)^2+
  \calE^{*2}(\calI-\calI')^2\right]\nonumber\\
  &&+{\textstyle\f{1}{8}}f_{19}\left[\calE\calE'(\calI^*-\calI'^*)^2+
  \calE^*\calE'^*(\calI-\calI')^2\right]+{\textstyle\f{1}{8}}f_{20}
  \left[\calE'^2(\calI^*-\calI'^*)^2+\calE'^{*2}(\calI-\calI')^2\right]
  \nonumber\\
  &&-{\textstyle\f{1}{8}}(2f_{19}+f_{13})
  \left(\calE\calE'^*-\calE^*\calE'\right)
  \left(\calI\calI'^*-\calI^*\calI'\right)\big\},
\end{eqnarray}
using coefficients $f_i$ as defined in Table~B.1 of \citet{MD99} with
$j=0$.  This expression can be shown to agree with that given by
\citet{LR95}.  Note that $R_{0,0}$ is real, and contributes negatively
to the secular Hamiltonian.

\subsection{Zeroth-order resonances}

Co-orbital $j:j$ resonance with $j\ge1$, for which $k=j$, $k'=-j$,
$\dot\phi_0=jn-jn'$ and $\alpha=1$:
\begin{eqnarray}
  \lefteqn{R=\f{GM'}{a'}\big\{(2f_1-\alpha\delta_{j,1})+
  (2f_2+{\textstyle\f{1}{2}}\alpha\delta_{j,1})
  \left(|\calE|^2+|\calE'|^2\right)+
  (2f_4-{\textstyle\f{1}{2}}f_2-{\textstyle\f{7}{64}}\alpha\delta_{j,1})
  |\calE|^4+(2f_5-{\textstyle\f{1}{4}}\alpha\delta_{j,1})|\calE|^2|\calE'|^2}&
  \nonumber\\
  &&+(2f_6-{\textstyle\f{1}{2}}f_2-{\textstyle\f{7}{64}}\alpha\delta_{j,1})
  |\calE'|^4+f_{17}\calE^2\calE'^{*2}+
  (\tilde f_{17}-{\textstyle\f{1}{64}}\alpha\delta_{j,1}-
  {\textstyle\f{81}{64}}\alpha\delta_{j,3})\calE^{*2}\calE'^2\nonumber\\
  &&+\calE\calE'^*\left[f_{10}+(f_{11}-{\textstyle\f{1}{8}}f_{10})|\calE|^2+
  (f_{12}-{\textstyle\f{1}{8}}f_{10})|\calE'|^2+{\textstyle\f{1}{4}}f_{13}
  \left(|\calI|^2+|\calI'|^2\right)+{\textstyle\f{1}{4}}f_{22}\calI^*\calI'+
  {\textstyle\f{1}{4}}f_{23}\calI\calI'^*\right]\nonumber\\
  &&+\calE^*\calE'\big[(\tilde f_{10}-\alpha\delta_{j,2})+(\tilde f_{11}-
  {\textstyle\f{1}{8}}\tilde f_{10}+{\textstyle\f{7}{8}}\alpha\delta_{j,2})
  |\calE|^2+(\tilde f_{12}-{\textstyle\f{1}{8}}\tilde f_{10}+
  {\textstyle\f{7}{8}}\alpha\delta_{j,2})|\calE'|^2+{\textstyle\f{1}{4}}
  (\tilde f_{13}+\alpha\delta_{j,2})\left(|\calI|^2+|\calI'|^2\right)
  \nonumber\\
  &&\quad+{\textstyle\f{1}{4}}\tilde f_{22}\calI\calI'^*+
  {\textstyle\f{1}{4}}(\tilde f_{23}-2\alpha\delta_{j,2})\calI^*\calI'\big]+
  {\textstyle\f{1}{4}}(2f_3+\alpha\delta_{j,1})
  \left(|\calI|^2+|\calI'|^2\right)+{\textstyle\f{1}{8}}f_8
  \left(|\calI|^4+|\calI'|^4\right)+{\textstyle\f{1}{16}}
  (2f_9-\alpha\delta_{j,1})|\calI|^2|\calI'|^2\nonumber\\
  &&+{\textstyle\f{1}{16}}f_{26}\calI^2\calI'^{*2}+{\textstyle\f{1}{16}}
  (\tilde f_{26}-\alpha\delta_{j,1})\calI^{*2}\calI'^2+{\textstyle\f{1}{4}}
  \calI\calI'^*\left[f_{14}+(f_{15}+{\textstyle\f{1}{4}}f_{14})
  \left(|\calE|^2+|\calE'|^2\right)+{\textstyle\f{1}{4}}f_{16}
  \left(|\calI|^2+|\calI'|^2\right)\right]\nonumber\\
  &&+{\textstyle\f{1}{4}}\calI^*\calI'
  \left[(\tilde f_{14}-2\alpha\delta_{j,1})+(\tilde f_{15}+
  {\textstyle\f{1}{4}}\tilde f_{14}+{\textstyle\f{1}{2}}\alpha\delta_{j,1})
  \left(|\calE|^2+|\calE'|^2\right)+{\textstyle\f{1}{4}}
  (\tilde f_{16}+\alpha\delta_{j,1})\left(|\calI|^2+|\calI'|^2\right)\right]
  \nonumber\\
  &&+{\textstyle\f{1}{4}}(f_3+2f_7)\left(|\calE|^2|\calI|^2+|\calE'|^2
  |\calI'|^2\right)+{\textstyle\f{1}{4}}(2f_7-{\textstyle\f{1}{2}}
  \alpha\delta_{j,1})\left(|\calE|^2|\calI'|^2+|\calE'|^2|\calI|^2\right)
  \nonumber\\
  &&+{\textstyle\f{1}{4}}(\calI^2+\calI'^2)\left(f_{18}\calE^{*2}+
  f_{19}\calE^*\calE'^*+f_{20}\calE'^{*2}\right)+{\textstyle\f{1}{4}}
  (\calI^{*2}+\calI'^{*2})\left(\tilde f_{18}\calE^2+\tilde f_{19}\calE\calE'+
  \tilde f_{20}\calE'^2\right)\nonumber\\
  &&+{\textstyle\f{1}{4}}\calI\calI'\left(f_{21}\calE^{*2}+
  f_{24}\calE^*\calE'^*+f_{25}\calE'^{*2}\right)+{\textstyle\f{1}{4}}
  \calI^*\calI'^*\left(\tilde f_{21}\calE^2+\tilde f_{24}\calE\calE'+
  \tilde f_{25}\calE'^2\right)\nonumber\\
  &&-{\textstyle\f{1}{32}}\alpha\delta_{j,1}\left[\calE^{*2}(\calI-\calI')^2+
  \calE'^2(\calI^*-\calI'^*)^2\right]\big\}.
\end{eqnarray}
Here $\tilde f_i$ denotes the coefficient $f_i$ in which $j$ is
replaced with $-j$.  Setting $\alpha=1$ renders the coefficients
singular, and this must be resolved in practice by a smoothing process
that represents averaging over the vertical extent of the disc.

\subsection{First-order resonances}

Interior $j:j-1:$ resonance with $j\ge2$, for which $k=j-1$, $k'=-j$,
$\dot\phi_0=(j-1)n-jn'$ and $\alpha=a/a'=[(j-1)/j]^{2/3}$:
\begin{eqnarray}
  \lefteqn{R=\f{GM'}{a'}\big\{f_{27}\calE+(f_{31}-2\alpha\delta_{j,2})\calE'+
  (f_{28}-{\textstyle\f{1}{8}}f_{27})|\calE|^2\calE+f_{29}|\calE'|^2\calE+
  (f_{32}+\alpha\delta_{j,2})|\calE|^2\calE'+(f_{33}-
  {\textstyle\f{1}{8}}f_{31}+{\textstyle\f{3}{2}}\alpha\delta_{j,2})|\calE'|^2
  \calE'}&\nonumber\\
  &&+f_{35}\calE^2\calE'^*+(f_{36}-{\textstyle\f{27}{16}}\alpha\delta_{j,3})
  \calE'^2\calE^*+{\textstyle\f{1}{8}}[f_{39}\calE+(f_{42}-4\alpha\delta_{j,2})
  \calE']\left[\calI^*(\calI'-\calI)-\calI'(\calI'^*-\calI^*)\right]
  \nonumber\\
  &&+{\textstyle\f{1}{8}}(f_{40}\calE+f_{43}\calE')\left[\calI(\calI'^*-
  \calI^*)-\calI'^*(\calI'-\calI)\right]+
  \f{1}{4}(f_{37}\calE^*+f_{38}\calE'^*)(\calI-\calI')^2\big\}.
\label{first-interior}
\end{eqnarray}
Exterior $j-1:j$ resonance with $j\ge2$, for which $k=-j$, $k'=j-1$,
$\dot\phi_0=-jn+(j-1)n'$ and $\alpha=a'/a=[(j-1)/j]^{2/3}$:
\begin{eqnarray}
  \lefteqn{R=\f{GM'}{a}\big\{f_{27}\calE'+(f_{31}-{\textstyle\f{1}{2}}
  \alpha^{-2}\delta_{j,2})\calE+(f_{28}-{\textstyle\f{1}{8}}f_{27})|\calE'|^2
  \calE'+f_{29}|\calE|^2\calE'+(f_{32}+{\textstyle\f{1}{4}}\alpha^{-2}
  \delta_{j,2})|\calE'|^2\calE}&\nonumber\\
  &&+(f_{33}-{\textstyle\f{1}{8}}f_{31}+{\textstyle\f{7}{16}}\alpha^{-2}
  \delta_{j,2})|\calE|^2\calE+f_{35}\calE'^2\calE^*+(f_{36}-
  {\textstyle\f{3}{4}}\alpha^{-2}\delta_{j,3})\calE^2\calE'^*\nonumber\\
  &&+{\textstyle\f{1}{8}}[f_{39}\calE'+(f_{42}-\alpha^{-2}\delta_{j,2})\calE]
  \left[\calI'^*(\calI-\calI')-\calI(\calI^*-\calI'^*)\right]
  \nonumber\\
  &&+{\textstyle\f{1}{8}}(f_{40}\calE'+f_{43}\calE)\left[\calI'(\calI^*-
  \calI'^*)-\calI^*(\calI-\calI')\right]+
  \f{1}{4}(f_{37}\calE'^*+f_{38}\calE^*)(\calI'-\calI)^2\big\}.
\end{eqnarray}

\subsection{Second-order resonances}

Interior $j:j-2$ resonance with $j\ge3$, for which $k=j-2$, $k'=-j$,
$\dot\phi_0=(j-2)n-jn'$ and $\alpha=a/a'=[(j-2)/j]^{2/3}$:
\begin{eqnarray}
  \lefteqn{R=\f{GM'}{a'}\big\{f_{45}\calE^2+f_{49}\calE\calE'+(f_{53}-
  {\textstyle\f{27}{8}}\alpha\delta_{j,3})\calE'^2+(f_{46}-
  {\textstyle\f{1}{4}}f_{45})|\calE|^2\calE^2+f_{47}|\calE'|^2\calE^2+
  (f_{50}-{\textstyle\f{1}{8}}f_{49})|\calE|^2\calE\calE'}&\nonumber\\
  &&+(f_{51}-{\textstyle\f{1}{8}}f_{49})|\calE'|^2\calE\calE'+(f_{54}+
  {\textstyle\f{27}{16}}\alpha\delta_{j,3})|\calE|^2\calE'^2+(f_{55}-
  {\textstyle\f{1}{4}}f_{53}+{\textstyle\f{135}{32}}\alpha\delta_{j,3})
  |\calE'|^2\calE'^2+f_{68}\calE^3\calE'^*+(f_{69}-{\textstyle\f{8}{3}}
  \alpha\delta_{j,4})\calE^*\calE'^3\nonumber\\
  &&+{\textstyle\f{1}{4}}f_{57}(\calI-\calI')^2+{\textstyle\f{1}{32}}f_{80}
  [\calI^3(\calI'^*-\calI^*)+\calI'^*(\calI^3-\calI'^3)]+{\textstyle\f{1}{32}}
  f_{81}[\calI'^3(\calI^*-\calI'^*)+\calI^*(\calI'^3-\calI^3)]\nonumber\\
  &&+{\textstyle\f{1}{64}}f_{61}[\calI^2\calI'(\calI'^*-\calI^*)-
  3\calI|\calI'|^2(\calI'-\calI)]+{\textstyle\f{1}{64}}f_{67}[\calI'^2
  \calI(\calI^*-\calI'^*)-3\calI'|\calI|^2(\calI-\calI')]\nonumber\\
  &&+{\textstyle\f{1}{8}}f_{57}(\calI-\calI')(|\calE|^2\calI-|\calE'|^2
  \calI')+{\textstyle\f{1}{8}}[f_{72}\calE^2+f_{74}\calE\calE'+(f_{78}-
  {\textstyle\f{27}{4}}\alpha\delta_{j,3})\calE'^2]\left[\calI^*
  (\calI'-\calI)-\calI'(\calI'^*-\calI^*)\right]\nonumber\\
  &&+{\textstyle\f{1}{8}}(f_{73}\calE^2+f_{75}\calE\calE'+f_{79}\calE'^2)
  \left[\calI(\calI'^*-\calI^*)-\calI'^*(\calI'-\calI)\right]+
  {\textstyle\f{1}{4}}(f_{58}|\calE|^2+f_{59}|\calE'|^2+f_{70}\calE^*\calE'+
  f_{71}\calE\calE'^*)(\calI-\calI')^2\big\}.
\label{second-interior}
\end{eqnarray}
%Could also make use of the relations $2f_{61}=-2f_{57}-5f_{80}-f_{81}$, $2f_{67}=-2f_{57}-f_{80}-5f_{81}$.}
Exterior $j-2:j$ resonance with $j\ge3$, for which $k=-j$, $k'=j-2$,
$\dot\phi_0=-jn+(j-2)n'$ and $\alpha=a'/a=[(j-2)/j]^{2/3}$:
\begin{eqnarray}
  \lefteqn{R=\f{GM'}{a}\big\{f_{45}\calE'^2+f_{49}\calE'\calE+(f_{53}-
  {\textstyle\f{3}{8}}\alpha^{-2}\delta_{j,3})\calE^2+(f_{46}-
  {\textstyle\f{1}{4}}f_{45})|\calE'|^2\calE'^2+f_{47}|\calE|^2\calE'^2+
  (f_{50}-{\textstyle\f{1}{8}}f_{49})|\calE'|^2\calE'\calE}&\nonumber\\
  &&+(f_{54}+{\textstyle\f{3}{16}}\alpha^{-2}\delta_{j,3})|\calE'|^2\calE^2+
  (f_{55}-{\textstyle\f{1}{4}}f_{53}+{\textstyle\f{15}{32}}\alpha^{-2}
  \delta_{j,3})|\calE|^2\calE^2+f_{68}\calE'^3\calE^*+(f_{69}-
  {\textstyle\f{2}{3}}\alpha^{-2}\delta_{j,4})\calE'^*\calE^3\nonumber\\
  &&+{\textstyle\f{1}{4}}f_{57}(\calI'-\calI)^2+{\textstyle\f{1}{32}}f_{80}
  [\calI'^3(\calI^*-\calI'^*)+\calI^*(\calI'^3-\calI^3)]+{\textstyle\f{1}{32}}
  f_{81}[\calI^3(\calI'^*-\calI^*)+\calI'^*(\calI^3-\calI'^3)]\nonumber\\
  &&+{\textstyle\f{1}{64}}f_{61}[\calI'^2\calI(\calI^*-\calI'^*)-3\calI'
  |\calI|^2(\calI-\calI')]+{\textstyle\f{1}{64}}f_{67}[\calI^2\calI'
  (\calI'^*-\calI^*)-3\calI|\calI'|^2(\calI'-\calI)]\nonumber\\
  &&+{\textstyle\f{1}{8}}f_{57}(\calI'-\calI)(|\calE'|^2\calI'-|\calE|^2
  \calI)+{\textstyle\f{1}{8}}[f_{72}\calE'^2+f_{74}\calE'\calE+(f_{78}-
  {\textstyle\f{3}{4}}\alpha^{-2}\delta_{j,3})\calE^2]\left[\calI'^*(\calI-
  \calI')-\calI(\calI^*-\calI'^*)\right]\nonumber\\
  &&+{\textstyle\f{1}{8}}(f_{73}\calE'^2+f_{75}\calE'\calE+f_{79}\calE^2)
  \left[\calI'(\calI^*-\calI'^*)-\calI^*(\calI-\calI')\right]+
  {\textstyle\f{1}{4}}(f_{58}|\calE'|^2+f_{59}|\calE|^2+f_{70}\calE'^*\calE+
  f_{71}\calE'\calE^*)(\calI'-\calI)^2\big\}.
\end{eqnarray}

\subsection{Third-order resonances}

Interior $j:j-3$ resonance with $j\ge4$, for which $k=j-3$, $k'=-j$,
$\dot\phi_0=(j-3)n-jn'$ and $\alpha=a/a'=[(j-3)/j]^{2/3}$:
\begin{equation}
  R=\f{GM'}{a'}\left[f_{82}\calE^3+f_{83}\calE^2\calE'+f_{84}\calE\calE'^2+
  (f_{85}-{\textstyle\f{16}{3}}\alpha\delta_{j,4})\calE'^3+
  {\textstyle\f{1}{4}}(f_{86}\calE+f_{87}\calE')(\calI-\calI')^2\right].
\end{equation}
Exterior $j-3:j$ resonance with $j\ge4$, for which $k=-j$, $k'=j-3$,
$\dot\phi_0=-jn+(j-3)n'$ and $\alpha=a'/a=[(j-3)/j]^{2/3}$:
\begin{equation}
  R=\f{GM'}{a}\left[f_{82}\calE'^3+f_{83}\calE'^2\calE+f_{84}\calE'\calE^2+
  (f_{85}-{\textstyle\f{1}{3}}\alpha^{-2}\delta_{j,4})\calE^3+
  {\textstyle\f{1}{4}}(f_{86}\calE'+f_{87}\calE)(\calI'-\calI)^2\right].
\end{equation}

\label{lastpage}


\begin{thebibliography}{}
  \bibitem[\protect\citeauthoryear{Artymowicz}{1993}]{A93}
    Artymowicz, P., 1993, ApJ 419, 155
  \bibitem[\protect\citeauthoryear{Borderies, Goldreich \& Tremaine}{1984}]
    {BGT84}
    Borderies, N., Goldreich, P., Tremaine, S., 1984, ApJ 284, 429
  \bibitem[\protect\citeauthoryear{Goldreich \& Sari}{2003}]{GS03}
    Goldreich, P., Sari, R., 2003, ApJ 585, 1024
  \bibitem[\protect\citeauthoryear{Goldreich \& Tremaine}{1978}]{GT78}
    Goldreich, P., Tremaine, S., 1978, ApJ 222, 850
  \bibitem[\protect\citeauthoryear{Goldreich \& Tremaine}{1979}]{GT79}
    Goldreich, P., Tremaine, S., 1979, ApJ 233, 857
  \bibitem[\protect\citeauthoryear{Goldreich \& Tremaine}{1980}]{GT80}
    Goldreich, P., Tremaine, S., 1980, ApJ 241, 425
  \bibitem[\protect\citeauthoryear{Goldreich \& Tremaine}{1981}]{GT81}
    Goldreich, P., Tremaine, S., 1981, ApJ 243, 1062
  \bibitem[\protect\citeauthoryear{Goodchild \& Ogilvie}{2006}]{GO06}
    Goodchild, S. G., Ogilvie, G. I., 2006, MNRAS 368, 1123
  \bibitem[\protect\citeauthoryear{Kne\v zevi\'c et~al.}{1991}]{KMFFF91}
    Kne\v zevi\'c, Z., Milani, A., Farinella, P., Froeschl\'e, Ch.,
    Froeschl\'e, Cl., 1991, Icarus 93, 316
  \bibitem[\protect\citeauthoryear{Laskar}{1997}]{L97}
    Laskar, J., 1997, A\&A 317, L75
  \bibitem[\protect\citeauthoryear{Laskar \& Robutel}{1995}]{LR95}
    Laskar, J., Robutel, P., 1995, Cel. Mech. Dynam. Astron. 62, 193
  \bibitem[\protect\citeauthoryear{Latter \& Ogilvie}{2006}]{LO06}
    Latter, H. N., Ogilvie, G. I., 2006, MNRAS, 372, 1829
  \bibitem[\protect\citeauthoryear{Lubow}{1991}]{L91}
    Lubow, S. H. 1991, ApJ 381, 259
  \bibitem[\protect\citeauthoryear{Lubow \& Ogilvie}{1998}]{LO98}
    Lubow, S. H., Ogilvie, G. I., 1998, ApJ 504, 983
  \bibitem[\protect\citeauthoryear{Lubow \& Ogilvie}{2001}]{LO01}
    Lubow, S. H., Ogilvie, G. I., 2001, ApJ 560, 997
  \bibitem[\protect\citeauthoryear{Meyer-Vernet \& Sicardy}{1987}]{MS87}
    Meyer-Vernet, N., Sicardy, B., 1987, Icarus 69, 157
  \bibitem[\protect\citeauthoryear{Morbidelli}{2002}]{M02}
    Morbidelli, A., 2002, Modern Celestial Mechanics, Taylor \& Francis,
    London
  \bibitem[\protect\citeauthoryear{Murray \& Dermott}{1999}]{MD99}
    Murray, C. D., Dermott, S. F., 1999, Solar System Dynamics,
    Cambridge Univ. Press, Cambridge
  \bibitem[\protect\citeauthoryear{Ogilvie \& Lubow}{2003}]{OL03}
    Ogilvie, G. I., Lubow, S. H., 2003, ApJ 587, 398
  \bibitem[\protect\citeauthoryear{Papaloizou}{2002}]{P02}
    Papaloizou, J. C. B., 2002, A\&A 388, 615
  \bibitem[\protect\citeauthoryear{Poincar\'e}{1892}]{P92}
    Poincar\'e, H., 1892, Les M\'ethodes Nouvelles de la M\'ecanique
    C\'eleste, vol.~1.  Gauthiers-Villars, Paris
  \bibitem[\protect\citeauthoryear{Strocchi}{1966}]{S66}
    Strocchi, F., 1966, Rev. Mod. Phys. 38, 36
  \bibitem[\protect\citeauthoryear{Ward}{1988}]{W88}
    Ward, W. R., 1988, Icarus 73, 330
  \bibitem[\protect\citeauthoryear{Ward}{1989}]{W89}
    Ward, W. R., 1989, ApJ 336, 526
  \bibitem[\protect\citeauthoryear{Yuan \& Cassen}{1994}]{YC94}
    Yuan, C., Cassen, P., 1994, ApJ 437, 338
  \end{thebibliography}
\end{document}